\documentclass[11pt]{article}
\usepackage{sharepackage}
\usepackage[margin=1in]{geometry}
\usepackage{xzhampackage}

\newif\ifanon
\anonfalse
\title{Two bases suffice for \( \QMAone \)-completeness}
\ifanon
\author{Anonymous}
\else
\author{Henry Ma\thanks{\texttt{henryma@mit.edu}} \\ MIT \and Anand Natarajan\thanks{\texttt{anandn@mit.edu}} \\ MIT}
\fi
\date{}

\begin{document}
\maketitle{}

\begin{abstract}
  We introduce a \textit{basis-restricted} variant of the \( \QkSAT \) problem, in which each term in the input Hamiltonian is required to be diagonal in either the standard or Hadamard basis. Our main result is that the \( \QsixSAT \) problem with this basis restriction is already \( \QMAone \)-complete, defined with respect to a natural gateset. Our construction is based on the Feynman-Kitaev circuit-to-Hamiltonian construction, with a modified clock encoding that interleaves two clocks in the standard and Hadamard bases. In light of the central role played by CSS codes and the uncertainty principle in the proof of the NLTS theorem of Anshu, Breuckmann, and Nirkhe (STOC '23), we hope that the CSS-like structure of our Hamiltonians will make them useful for progress towards a quantum PCP theorem.
\end{abstract}

\section{Introduction}

Local Hamiltonians play a central role in quantum complexity theory, tying the subject both to applications in condensed matter physics and quantum chemistry, and to the classical theory of $\NP$-completeness, constraint satisfaction problems, and combinatorial optimization. Much of the complexity theory of local Hamiltonians has developed in analogy to the theory of $\NP$-completeness, with the class $\QMA$ playing the role of $\NP$, and Kitaev's result showing that the \emph{local Hamiltonian problem} (estimating the ground energy of a local Hamiltonian) is $\QMA$-complete playing the role of the Cook-Levin theorem ($\NP$-completeness of 3SAT, and local constraint satisfaction problems more generally). Despite these analogies, there are considerable differences between the classical and quantum settings. One important difference relates to the issue of ``perfect completeness.'' In the case of classical constraint satisfaction problems, it is typically just as hard to solve the ``SAT'' problem of deciding whether all constraints are simultaneously satisfiable, as to solve the ``decisional MAX-SAT'' problem of deciding whether at least a certain fraction of the constraints are satisfiable: both problems are $\NP$-complete. In contrast, the equivalent SAT and MAX-SAT problems for local Hamiltonians appear to have different complexities: all of our known $\QMA$-completeness results are MAX-SAT type results, while SAT-type problems appear to capture the complexity class $\QMAone$ of $\QMA$ proof systems with \emph{perfect completeness} (i.e. the verifier accepts good proofs with certainty). By definition, $\QMAone \subseteq \QMA$, and the two classes are believed to be qualitatively similar in power, but the exact relation between them is murky.

Due to the nature of our tools for showing computational hardness of local Hamiltonian problems, our understanding of the MAX-SAT setting is better than our understanding of the SAT setting: in the MAX-SAT world, we have dichotomy theorems~\cite{CM16ComplexityClassificationLocal} for 2-local Hamiltonians, and a wide variety of families of Hamiltonians are known to be $\QMA$-complete, including very simple and natural models like the $XZ$-model~\cite{BL08RealizableHamiltoniansUniversal}. In the SAT setting, while we know that $\QthreeSAT$ (the SAT problem for 3-local Hamiltonians) is $\QMAone$-complete~\cite{GN16Quantum3SATQMA1Complete}, and that various other cases of $\QkSAT$ are easier (e.g. for $k=2$ the problem is in $\P$~\cite{Bra06EfficientAlgorithmQuantum}, and for so-called ``stoquastic'' Hamiltonians the problem is in $\ccs{MA}$~\cite{bravyi2010complexity}), we don't have dichotomy theorems, nor do we have hardness results as clean as the $XZ$-model hardness mentioned above. This is because tools like \emph{gadget reductions} used to reduce between different types of Hamiltonians do not preserve perfect completeness. 

At the same time, the class $\QMAone$ is scientifically of great interest as a testbed to better understand Hamiltonian complexity. Arguably the single most important open problem in Hamiltonian complexity is the quantum PCP conjecture: that approximating the ground energy of a local Hamiltonian even up to a constant fraction of the total norm of the Hamiltonian remains $\QMA$-hard. A natural weakening of this conjecture would be to show $\QMAone$-hardness for this problem, and in fact, the biggest partial milestone we have towards a quantum PCP theorem comes from the SAT setting. Specifically, the NLTS theorem of Anshu, Breuckmann, and Nirkhe~\cite{ABN23} shows the existence of local Hamiltonians whose low-energy states satisfy a certain structural property that we expect should hold for quantum PCP Hamiltonians. These Hamiltonians are parent Hamiltonians of quantum stabilizer error correcting codes, and thus are exactly satisfiable (all valid code states are $0$-energy ground states)---and this characterization of the ground space is used explicitly in the proof of the NLTS theorem. However, for the same reason, these Hamiltonians contain no computational hardness. The most natural next step beyond the NLTS theorem would be to show that the NLTS structural property holds for a family of Hamiltonians where the local Hamiltonian problem is \emph{computationally hard}. In order to achieve such a result, can we show a $\QMAone$-hardness result for a class of local Hamiltonians that is sufficiently structured to use the techniques of Anshu, Breuckmann and Nirkhe's proof?

In this work, we focus on one structural property of the code Hamiltonians of~\cite{ABN23}: every local term in the Hamiltonian is a projector and is diagonal in either the standard basis (the ``Pauli $Z$-basis''), or the Hadamard basis (the ``Pauli $X$-basis''). Stabilizer codes with this property are called \emph{CSS codes}, and the CSS structure of the code is crucial to the proof of the NLTS theorem in two ways. Firstly, to prove the NLTS property of low-energy states, Anshu et al, following the approach of Eldar and Harrow~\cite{EH17LocalHamiltoniansWhose}, reduce the problem to proving that low-energy states of their Hamiltonian, when measured in either the standard or Hadamard bases, give rise to ``well spread'' probability distributions over the Boolean hypercube. They are able to prove this property in turn using a Heisenberg uncertainty principle relating these two bases, together with properties of the codes they consider. Secondly, these code properties in turn are formulated and proved by viewing CSS codes as consisting of a pair of \emph{classical} error correcting codes, one in each basis. In fact, the view of CSS codes as chain complexes, which is at the foundation of the vast literature on constructing and analyzing such codes, relies on this two-basis structure.

The main result of this work is to show that the SAT problem for two-basis local Hamiltonians is $\QMAone$-complete. More precisely, our main theorem is the following.
\begin{thm}
  \label{thm:main}
  Let $\XZQsixSAT$ be the following problem: given a local Hamiltonian $H = \sum_i H_i$ on $n$ qubits, where each local term $H_i$ is a projector acting on $6$ qubits and is diagonal in either the standard or Hadamard basis, determine whether $\lambda_{\min}(H) = 0$ or $\lambda_{\min}(H) \geq b(n)$ where $b(n) = \Omega(1/\poly(n))$, promised that one of the two is the case. Then for an appropriately chosen function $b(n)$, it holds that $\XZQsixSAT$ is complete for the class $\QMAone^{\gG_2}$ of quantum Merlin-Arthur proof systems with perfect completeness where the verifier consists of a circuit made up of gates from the gate set $\gG_2$. 
\end{thm}
There is a technical subtlety in the theorem statement, arising from the dependence of $\QMAone$ on the choice of gate set for the verifier's circuit. We show completeness with respect to the gate set $\gG_2$ as defined by Rudolph \cite{Rud25UniversalGatesetQMA1}, consisting of NOT, controlled-NOT, and Toffoli gates ($X, CX, CCX$) and the tensor product $\hat{H} \otimes \hat{H}$ of two Hadamard gates. This gate set is a slight variant of the commonly used Hadamard-plus-Toffoli gate set~\cite{Aha03SimpleProofThat}, which is universal for quantum computing.

To better understand the class of Hamiltonians for which we show hardness, it is perhaps useful to draw a classical analogy.  A classical \emph{linear} code is associated with a collection of linear constraints over $\mathbb{F}_2$. By the Gaussian elimination algorithm, the SAT problem for linear constraints can always be solved in polynomial time, but once the constraints are allowed to be nonlinear, the complexity of the problem jumps up to $\NP$. In the quantum world, the parent Hamiltonian of a CSS code consists of a pair of linear classical constraint satisfaction problems (CSPs), one for each basis: a valid code state is one that, when measured in the $Z$-basis, yields a satisfying string for the $Z$-basis linear constraints,  and when measured in the $X$-basis, yields a satisfying string for the $X$-basis linear constraints. By standard stabilizer techniques, Gaussian elimination is sufficient to solve the SAT problem for these Hamiltonians in polynomial time. The Hamiltonians we consider also consist of a pair of classical constraint satisfaction problems in the two bases, except with nonlinear\footnote{To be clear, the constraints are nonlinear in their action on the binary string measurement outcomes viewed as elements of $\mathbb{F}_2^n$---the Hamiltonian is still a linear operator over the Hilbert space.} constraints now being allowed, and our result shows that, as in the classical case, allowing for nonlinearity causes the complexity to jump from $\P$ to the maximal possible level of hardness ($\QMAone$ in this case). This classical analogy also helps illustrate the relation between our result and the MAX-SAT case. In the MAX-SAT case, classically, even two-local linear constraints become $\NP$-hard (this is by the $\NP$-hardness of the MAX-CUT problem), and similarly, in the quantum case, the result of Biamonte and Love for the $XZ$-model shows that the MAX-SAT problem for ``linear'' constraints is $\QMA$-hard.

\begin{table}
\centering
\begin{tabular}{c | c | c}
  Constraints & SAT & MAX-SAT  \\ 
  \hline
  $Z$-basis, linear  & In $\P$ (by Gaussian elimination) & $\NP$-hard (by MAX-CUT) \\
  $Z$-basis, general & $\NP$-hard (Cook-Levin) & $\NP$-hard (by $\leftarrow$) \\
  $Z$ and $X$ bases, linear & In $\P$ (by Gaussian elimination) & $\QMA$-hard~\cite{BL08RealizableHamiltoniansUniversal} \\
  $Z$ and $X$ bases, general & \textcolor{purple}{$\QMAone$-hard (this work)} & $\QMA$-hard (by $\uparrow$)
\end{tabular}
\end{table}

\paragraph{The two basis paradigm.}
    Viewing matters more subjectively and at a higher level, it is a striking fact that many interesting phenomena in quantum computing rest on the interplay between the standard and Hadamard bases. Examples of this ``two basis paradigm'' in action include the BB84 protocol and Wiesner's quantum money scheme, Simon's algorithm, the magic square game (and, arguably, the proof that $\ccs{MIP^*} = \ccs{RE}$), the forrelation problem (some versions of which are $\ccs{BQP}$-complete), Aaronson and Christiano's subspace quantum money scheme, and Mahadev's measurement protocol, besides the previously mentioned examples of CSS codes and Biamonte and Love's QMA-hardness results. We see our result as another instance of this paradigm.

\subsection{Technical overview}
Our proof of \Cref{thm:main} consists of two parts. To show $\QMAone^{\gG_2}$-completeness of our problem, we must show that it is both contained in $\QMAone^{\gG_2}$ and that it is $\QMAone^{\gG_2}$-hard. The containment follows along standard lines: we construct a verifier for $\XZQsixSAT$ instances that samples a random term from the Hamiltonian and coherently measures it on the witness state. The only nontrivial step in this construction is showing that the verifier's gate set $\gG_2$ can exactly simulate a controlled Hadamard gate, since this gate is needed to perform the measurement of the witness in the appropriate basis.

The bulk of the proof is concerned with the $\QMAone^{\gG_2}$-hardness. We build on the Kitaev circuit-to-Hamiltonian construction \cite{KSV02ClassicalQuantumComputation},
which is a quantum analog of the fundamental construction used by Cook and Levin \cite{Coo71ComplexityTheoremprovingProcedures, Lev73UniversalSequentialSearch} to show that \( \SAT \) is \( \NP \)-complete. Let us consider a circuit consisting of gates $U_1, U_2, \dots, U_T$, acting on a Hilbert space $\mathcal{H}$. Kitaev's construction associates this circuit to a Hamiltonian $H$ acting on a space $\mathcal{H} \otimes \mathcal{H}_{\mathrm{clock}}$, where $\mathcal{H}_{\mathrm{clock}}$ is the Hilbert space of an ancillary ``clock'' register. The space $\mathcal{H}_{\mathrm{clock}}$ contains at least $T+1$ orthonormal states $\ket{\hat{0}}, \ket{\hat{1}}, \dots, \ket{\hat{T}}$, but typically has a much larger dimension---$2^T$ in Kitaev's original construction. The full Hamiltonian $H$ can be written as a sum of four components:

\[ H = H_{\mathrm{prop}} + H_{\mathrm{format}} + H_{\mathrm{in}} + H_{\mathrm{out}},\]
where each component is a sum of local projectors. Moreover, Kitaev's analysis shows that, when applied to a $\QMAone$ circuit $U_1, \dots, U_T$ that accepts some witness state with certainty, the Hamiltonian $H$ has a $0$- energy ground state, and otherwise, if applied to a circuit that rejects all witnesses with high probability, then the minimum energy of $H$ is at least $1/\poly(n)$. 

Our approach is to use Kitaev's construction, but modify the terms $H_{\mathrm{prop}}$ and $H_{\mathrm{format}}$, as well as the encoding of the clock states $\ket{\hat{0}}, \dots, \ket{\hat{T}}$, so that the resulting Hamiltonian $H$ satisfies our two-basis constraint. Since we will not modify them significantly, for now let us ignore the terms $H_{\mathrm{init}}$ and $H_{\mathrm{out}}$, and study the ground space of the remaining two terms. These are designed in Kitaev's construction so that they \emph{always} have a nonempty 0-energy ground space, which consists of \emph{history states} of the form
\[ \ket{\psi_{\mathrm{history}}} = \frac{1}{\sqrt{T+1}} \sum_{t=0}^{T} U_t \dots U_1 \ket{\psi_0} \otimes \ket{\hat{t}}.\]
This is achieved in two ways:
\begin{itemize}
\item The ``format'' Hamiltonian $H_{\mathrm{format}}$ forces the state to be supported only on the ``good'' clock register states $\ket{\hat{0}}, \dots, \ket{\hat{T}}$. This is done in Kitaev's construction by taking $\mathcal{H}_{\mathrm{clock}}$ to be a space of $T$ qubits, and taking the good clock states to be encodings of $0, \dots, T$ in unary in the standard basis, so that
    \[ \ket{\hat{t}} = \lket{1}_1 \dots \lket{1}_{t-1} \lket{1}_t \lket{0}_{t+1} \dots \lket{0}_T. \]
    To force the state to be supported only on these states, $H_{\mathrm{format}}$ consist of projectors that enforce the constraint on each pair of adjacent qubits $(t,t+1)$ that a $0$ can never be followed by a $1$. These constraints are already diagonal in the $Z$-basis.
    \item The ``propagation'' Hamiltonian $H_{\mathrm{prop}}$ consists of a sum of $T$ constraints, each of which forces the $\ket{\widehat{t-1}}$ and $\ket{\hat{t}}$ components of the state to be related by an application of the unitary $U_{t}$. In Kitaev's construction, these terms take the form
    \begin{align*} H_{\mathrm{prop},t} &= \frac{1}{2} I \otimes (\lktbr{110}{110} + \lktbr{100}{100})_{t-1,t,t+1} \\
    &\qquad - \frac{1}{2} U_t \otimes (\lktbr{110}{100})_{t-1,t,t+1} - \frac{1}{2} U_t^\dagger \otimes (\lktbr{100}{110})_{t-1,t,t+1}, \end{align*}
    where the second tensor factor acts on the specified qubits of the clock register.
    It can be checked that for general choices of $U_t$, these terms are not diagonal in either the $X$-basis or the $Z$-basis.
\end{itemize}
To make progress, we must modify the propagation terms. To do this, we make the following key observation: the terms $H_{\mathrm{prop},t}$ are ``almost'' diagonal in the $X$-basis whenever the unitary $U_t$ itself is diagonal in the $X$-basis and is Hermitian (i.e. $U_t = U_t^\dagger$). Specifically, in this case, $H_{\mathrm{prop},t}$ can be written as
\[ H_{\mathrm{prop},t} = \frac{1}{2} (I \otimes I_t - U_t \otimes X_t) \otimes (\lktbr{10}{10})_{t-1,t+1},\]
which is diagonal when all qubits are placed in the $X$-basis except for qubits $t-1, t+1$ of the clock register, which remain in the $Z$-basis. Moreover, the same property would hold, with $X$ and $Z$ interchanged, if the good clock states were $X$-basis states, and $U_t$ were Hermitian and diagonal in the $Z$-basis.

This suggests the following modifications:
\begin{itemize}
    \item Encode good clock states by putting alternating qubits in the $X$- and $Z$-bases. This means that valid clock states would look like
    \begin{align*}
        \ket{\hat{0}} &= \ket{\clockfont{0+0+0+}\dots} \\
        \ket{\hat{1}} &= \ket{\clockfont{1+0+0+}\dots} \\
        \ket{\hat{2}} &= \ket{\clockfont{1-0+0+}\dots} \\
        \ket{\hat{3}} &= \ket{\clockfont{1-1+0+}\dots} \\
        &\dots. 
    \end{align*}
    \item Choose a universal gate set where each gate is Hermitian and diagonal in either the $X$- or $Z$-basis. We construct such a gate set and show that it is computationally equivalent to $\gG_2$.
    \item Padding the circuit so that each $X$-basis gate falls on an odd timestep and each $Z$-basis timestep falls on an even timestep. This makes the propagation terms fully diagonal in the $X$-basis for odd times, and the $Z$-basis for even times: e.g. for odd $t$,
    \[ H_{\mathrm{prop}, t} = \frac{1}{2} (I \otimes I_t - U_t \otimes X_t) \otimes \lktbr{-+}{-+}_{t-1,t+1}. \]
\end{itemize}
With these changes, every term in $H_{\mathrm{prop},t}$ is diagonal in either the $X$- or $Z$-basis. However, now that we have changed the encoding of the clock, we must also change $H_{\mathrm{format}}$, and it is not hard to see that any Hamiltonian that restricts us only to valid clock states will not be diagonal in either basis. The solution to this is somewhat surprising.
\begin{itemize}
    \item We modify $H_{\mathrm{format}}$ to include checks that only check consistency \emph{separately} on the $X$- and $Z$-basis parts of the clock state. Specifically, our new $H_{\mathrm{format}}$ checks that the clock register in a state $\ket{t_Z, t_X}$ consisting of an ``interleaving'' of some valid unary-encoding of an integer $t_Z$ in the $Z$-basis in odd positions, and an integer $t_X$ in the $X$-basis in even positions. It does \emph{not} check that these two interleaved times are synchronized with each other, meaning that this Hamiltonian accepts invalid sates such as
    \[ \ket{\clockfont{0-0-0+}\dots\clockfont{0+0}},\]
    which do not correspond to valid clock states. We call these states ``fake'' states.
    \item Surprisingly, we show that, in fact, $H_{\mathrm{prop}} + H_{\mathrm{format}}$ \emph{together} force the ground state to supported only on valid clock states. This is because every fake clock state is coupled by some term in $H_{\mathrm{prop}}$ to a ``bad'' clock state, which incurs an energy penalty from $H_{\mathrm{fromat}}$. We analyze this quantitatively by relating the Hamiltonian to a graph Laplacian. 
\end{itemize}
The majority of the technical work in our analysis consists in (a) proving universality of our new gate set, and (b) quantitatively bounding the minimum energy of $H_{\mathrm{format}} + H_{\mathrm{prop}}$ on the subspace of invalid clock strings.

\subsection{Open questions}
We see several possibilities for future work with a quantum PCP flavor.
\begin{enumerate}
\item Our construction, since it is based on the Feynman-Kitaev clock, cannot achieve the NLTS property: indeed, the product state $\ket{0} \otimes \ket{\hat{0}}$ violates only a single propagation term of the Hamiltonian. To get to NLTS while maintaining the two-basis structure, can we apply our ideas to the tensor-network-based circuit-to-Hamiltonian construction of~\cite{anshu2024circuit}?
\item What do random instances of our two-basis Hamiltonians look like? Can we find a distribution such that Hamiltonians sampled from this distribution have a zero-energy ground state with high probability, but where finding the ground state is computationally hard? It would be especially interesting to relate this to work on the combinatorial NLTS property for tensor network Hamiltonians constructed from random SAT instances~\cite{AGK23}.
\item One candidate approach to proving the quantum PCP conjecture is to design a ``locality-preserving gap amplification'' procedure, that operates on instances of local Hamiltonians by increasing the minimum energy of unsatisfiable Hamitonians (amplifying the ``promise gap''), while preserving the locality of the Hamiltonian. To build towards such a procedure, can we apply classical ``gap-amplification'' or ``locality-reduction'' transformations separately to one of the two bases and make some kind of meaningful progress towards gap amplification? Moreover, is there an analog of ``distance balancing'' for CSS codes~\cite{wills2023general}, whereby a code with high distance in one basis but low distance in the other can be converted into a code with decent distance in both bases?
\item In this paper we have taken the point of view that $\QMAone^{\gG_2}$ is likely to be qualitatively similar to $\QMA$ in power. But if $\QMAone^{\gG_2}$ is in fact much weaker, could our completeness result give us a route to showing this by putting the problem $\XZQsixSAT$ into a smaller class like $\ccs{QCMA}$? One very speculative route to doing this is to give an efficient classical description of ground states of such Hamiltonians, perhaps using tools from additive combinatorics, since Hamiltonian terms in the $X$-basis can be viewed as additive constraints on the support of the ground state in the standard basis.
\end{enumerate}

\ifanon
%blank acknowledgements section
\else
\paragraph{Acknowledgements.} Part of this work was done while both authors were visiting the Simons Institute for the Theory of Computing as part of the 2025 Summer Cluster on Quantum Computing, and the Challenge Institute for Quantum Computation at UC Berkeley. HM was supported by the NSF Graduate Research Fellowship Program. AN was supported by NSF CAREER grant number 2339948. We thank Chinmay Nirkhe for several helpful discussions.
\fi

\section{Preliminaries}

\subsection{Quantum computation}

We briefly set up the basic formalism of quantum computation.
For a more detailed exposition, we refer the reader to \cite{NC10QuantumComputationQuantum}.

An \( n \)-qubit \emph{quantum state} \( \ket{\psi} \) is a unit vector in a complex Hilbert space \((\C^2)^{\otimes n} \).
A \( k \)-qubit \emph{quantum operation} is a unitary operator \( U \) acting on a \( k \)-qubit space \( (\C^2)^{\otimes k} \).
We follow the conventions of Dirac notation.
We use \( U^\dagger \) to denote the adjoint of an operator.
Let \( [n] = \set{1, \dots, n} \).
We can extend \( U \) to an operation \( U \otimes I_{[n] \setminus S} \) on \( n \)-qubit space, where \( S \subseteq [n]\) is the set of \( k \) qubits which \( U \) acts on, and \( I_{[n] \setminus S} \) is the identity operator on the remaining qubits.

We now fix some notation for relevant states and operations.
As usual, \( \ket{0} \) and \( \ket{1} \) refer to the single-qubit standard basis states, while \( \ket{+} \) and \( \ket{-} \) are the Hadamard basis states.
We will often omit tensor products, e.g. \( \ket{00} = \ket{0} \otimes \ket{0} \).
Let \( X \) and \( Z \) be the corresponding single-qubit Pauli operations.
We will also refer to the standard and Hadamard bases as the \( Z \) basis and \( X \) basis respectively.
Let \( \had \) be the Hadamard gate (we put the hat to avoid confusion with a Hamiltonian \( H \)).
Let \( CX \), \( CZ \), and \( \swap \) denote the two-qubit CNOT, controlled-\( Z \) and swap operations respectively.
Let \( CCX \) denote the Toffoli gate and \( CCZ \) the controlled-controlled-\( Z \) gate.

A \emph{quantum circuit} on \( n \) qubits is an \( n \)-qubit operation \( U \) which can be decomposed into a sequence of gates \( U = G_m \cdots G_1 \),
where each \( G_i \) is a quantum operation from some fixed \emph{gate set}; typically, this gate set contains operations which each acts nontrivially on just a small number of qubits.
In our protocols, we will consider projective measurements of the first qubit in the \( Z \) basis, which is given by the orthogonal projectors \( \set{\ktbr{0}{0}_1 \otimes I, \ktbr{1}{1}_1 \otimes I} \),
where the subscript indicates the first qubit register, and the identity operation acts on the remaining qubits.

\subsection{Quantum complexity theory}

We now define the model of quantum verification of interest in this work.
A \emph{promise problem} is a pair \( (L_{\mathrm{yes}}, L_{\mathrm{no}}) \) of subsets of bitstrings \( L_{\mathrm{yes}}, L_{\mathrm{no}} \subseteq \B^\star \) which satisfies \( L_{\mathrm{yes}} \cap L_{\mathrm{no}} = \varnothing \).
Promise problems generalize languages, a notion from classical complexity theory: a language is just a promise problem which additionally satisfies \( L_{\mathrm{yes}} \cup L_{\mathrm{no}} = \B^\star \).

A \emph{quantum verifier} is a uniform family of quantum circuits \( \set{V_x} \) whose goal is to determine whether a given string \( x \) is in \( L_{\mathrm{yes}} \) or \( L_{\mathrm{no}} \), promised that one is the case.
On an input \( x \) of length \( n \), the verifier is given access to a quantum proof state \( \ket{\psi} \) and a register of ancilla qubits all initialized to \( \ket{0} \).
The circuit is efficient, in the sense that both the proof and ancilla registers have \( \poly(n) \) qubits,
and there are \( \poly(n) \) gates in the circuit.

After the circuit is applied, the final step of the verification is to measure the first qubit in the \( Z \) basis.
We say the verifier accepts if the measurement outcome is \( 1 \) and rejects if the outcome is \( 0 \).
The probability that the verifier \( V_x \) accepts is then
\begin{align}
\label{eq:1}
\norm{(\ktbr{1}{1}_1 \otimes I)V_x\ket{\psi}\ket{0 \cdots 0}_{\text{anc}}}^2.
\end{align}

We now formally define the classes \( \QMAone \) and \( \QMA \); we will mainly be interested in \( \QMAone \) in this work.
\begin{defn}
  A promise problem \( (L_{\mathrm{yes}}, L_{\mathrm{no}}) \) is in \( \QMAone \) if there is a uniform family of efficient quantum circuits \( \set{V_x} \) such that for any \( x \in \B^n \),
  \begin{enumerate}
  \item If \( x \in L_{\mathrm{yes}} \), then there is some proof state for which \( V_x \) accepts with probability \( 1 \).
    \item If \( x \in L_{\mathrm{no}} \), then for any proof state, \( V_x \) accepts with probability \( \le 1 / 3 \).
  \end{enumerate}
\end{defn}
We refer to the first condition as \emph{completeness} and the second condition as \emph{soundness}.
Importantly, the definition of \( \QMAone \) is not known to be independent of the choice of gate set for the circuits.
In this work, we choose the gate set
\[ \gs = \set{X, CX, C C X, \had \otimes \had} \] for \( \QMAone \).
When needing to explicitly refer to \( \QMAone \) with this gate set (e.g. in comparison with other gate sets),
we use the notation \( \QMAone^{\gs} \).

We make some observations about the chosen gate set.
\begin{rmk}
First, \( \gs \) is computationally universal, i.e. any quantum circuit can be simulated by a circuit which only uses gates from \( \gs \).
This is because gates in \( \gs \) allow us to implement \( C C X \) and \( \had \) (which can be implemented by applying \( \had \otimes \had \) on the desired qubit and an otherwise unused ancilla qubit); these two gates already form a computationally universal gate set \cite{Shi03BothToffoliControlledNOT, Aha03SimpleProofThat}.

Second, \( \gs \) is studied in a work of Rudolph \cite{Rud25UniversalGatesetQMA1} which makes progress on the interesting open problem of whether \( \QMAone \) has a universal gate set.
Among other results related to \( \gs \), they show that the problem of Gapped Clique Homology on weighted graphs (introduced in \cite{KK24GappedCliqueHomology}) is \( \QMAone^{\gs}\)-complete.
This result gives a surprising connection between an important problem in computational topology and \( \QMAone^{\gs} \), motivating our study of other properties of \( \QMAone^{\gs} \).
\end{rmk}

The class \( \QMA \) is defined in the same way as \( \QMAone \), except the completeness parameter is \( 2 / 3 \),
i.e. for \( x \in L_{\mathrm{yes}} \), the verifier must accept with probability at least \( 2 / 3 \).
In contrast to \( \QMAone \), the definition of \( \QMA \) is independent of the choice of gate set, since the Solovay-Kitaev theorem \cite{DN06SolovayKitaevAlgorithm} shows that any quantum gate can be efficiently approximated using gates from a universal gate set.
This argument does not straightforwardly extend to \( \QMAone \): approximation of a gate is not enough, since this may not preserve the perfect completeness required of a \( \QMAone \) protocol.

The classical theory of \( \NP \)-completeness generalizes to \( \QMAone \) and \( \QMA \) in a straightforward way.
A promise problem \( K = (K_{\mathrm{yes}}, K_{\mathrm{no}}) \) has an \emph{efficient reduction} to a promise problem \( L = (L_{\mathrm{yes}}, L_{\mathrm{no}}) \) if there is an efficient deterministic algorithm \( A \) which maps a bitstring \( x \) to a bitstring \( A(x) \) such that \( A(x) \in L_{\mathrm{yes}} \) if \( x \in K_{\mathrm{yes}} \), and \( A(x) \in L_{\mathrm{no}} \) if \( x \in K_{\mathrm{no}} \).
A promise problem \( L \) is \emph{\( \QMAone \)-hard} if for every \( K \in \QMAone \), there is an efficient reduction from \( K \) to \( L \).
A promise problem \( L \) is \emph{\( \QMAone \)-complete} if \( L \) is in \( \QMAone \) and \( L \) is \( \QMAone \)-hard.
The definitions for \( \QMA \) are analogous.

\subsection{Hamiltonian complexity}

In this section, we define the \( \QkSAT \) problem and introduce its basis-restricted variant.
An \( n \)-qubit \emph{Hamiltonian} is a Hermitian operator acting on \( n \) qubits.
A \emph{\( k \)-local operator} on \( n \) qubits is an operator which acts nontrivially on at most \( k \) of the qubits.
The \emph{energy} of a state \( \ket{\psi} \) with respect to Hamiltonian \( H \) is \( \bra{\psi} H \ket{\psi} \).
Note that this quantity is always real and non-negative, since \( H \) is Hermitian.

\begin{defn}
  Let \( k \ge 1 \) and let \( \sS \) be a set of Hermitian \( k \)-local projectors.
  \( \QkSAT \) is a promise problem whose input is a classical description of a Hamiltonian \( H = \sum_{i = 1}^m h_i \),
  where each term \( h_i \) acts on \( n \) qubits and belongs to \( \sS \).
  \( H \) is a ``yes'' instance if there is an \( n \)-qubit state \( \ket{\psi} \) such that \( \bra{\psi} H \ket{\psi} = 0 \).
  \( H \) is a ``no'' instance if for all \( n \)-qubit states \( \ket{\psi} \), \( \bra{\psi} H \ket{\psi} \ge 1 / \poly(n) \).
\end{defn}
Note that the definition of \( \QkSAT \) implicitly depends on the choice of allowed projectors \( \sS \); this will be relevant in defining its basis-restricted variant.

As an illustrating example,  we first recall how \( \QkSAT \) generalizes the \( \NP \)-complete problem \( \kSAT \).
For simplicity we consider the case \( k = 3 \): let \( \varphi = C_1 \land \cdots \land C_m \) be a \( \threeSAT \) instance on \( n \) variables \( x_1, \dots, x_n \).
We construct an instance \( H = \sum_{i=1}^m h_i \) of \( \QthreeSAT \).
Recall that clause \( C \) has the form \( C = l_1 \lor l_2 \lor l_3 \) for some literals \( l_j \).
Set \( a_j \in [n] \) so that \( l_j \) is the variable \( x_{a_j} \) or its negation (we can assume that the \( a_j \)'s are distinct).
Let \( z = z_1z_2z_3 \in \B^3 \) be the unique assignment to \( l_1 \), \( l_2 \), and \( l_3 \) which does not satisfy \( C \).
Define a \( 3 \)-local \( n \)-qubit projector \( h \) corresponding to \( C \) by \( h = \ktbr{z}{z}_{a_1,a_2,a_3} \otimes I \),
i.e. \( h \) acts as the projector \( \ktbr{z_j}{z_j} \) on qubit \( a_j \) for \( j = 1,2,3\),
and acts trivally on the remaining qubits.
\( H \) is constructed by defining each \( h_i \) from clause \( C_i \) in this manner.

For any assignment \( a \in \B^n \) to \( \varphi \),
if \( a \) satisfies \( C_i \) then \( \bra{a} h_i \ket{a} = 0 \), and otherwise \( \bra{a}  h_i \ket{a} = 1\).
Thus, If \( \varphi \) is satisfiable with some assignment \( a \), the state \( \ket{a} \) is a zero-energy state of \( H \).
If \( \varphi \) is not satisfiable, then any \( Z \) basis state \( \ket{z} \) satisfies \( \bra{z} H \ket{z} \ge 1 \),
and thus \( \bra{\psi} H \ket{\psi} \ge 1 \) for any \( n \)-qubit state \( \ket{\psi} \).
This shows how \( \threeSAT \) can be viewed as a special case of \( \QthreeSAT \).

Note that the Hamiltonian which arises from this transformation take on a highly restricted form:
each projector term is diagonal in the same, predetermined basis (namely, the \( Z \) basis).
Let us refer to this as a \emph{classical Hamiltonian}.
This brings into view one of the guiding questions of this work:
if we relax the restrictions placed on a classical Hamiltonian, at what point does the Hamiltonian become ``quantum''?
This question has been previously studied for the relaxation in which the Hamiltonian is no longer required to be classical,
but instead the Hamiltonian terms must pairwise commute.
In this work, we introduce a new way of generalizing classical Hamiltonians which we call \emph{basis restriction}.

\begin{defn}
  For \( n \ge 1 \), let \( \bB_n \) be a set of bases of \( n \)-qubit space.
  Let \( \bB = \cup_{n \ge 1} \bB_n \).
  Define \( \bB \)-\( \QkSAT \) to be the problem \( \QkSAT \) where we choose \( \sS \) (the set of allowed projectors) to be
  the set of Hermitian \( k \)-local projectors which are diagonal in some basis in \( \bB \).
\end{defn}

As an example, if for all \( n \) we let \( \bB_n \) include just the \( Z \) basis on \( n \) qubits,
then an instance of \( \bB \)-\( \QkSAT \) is a Hamiltonian \( H = \sum_i h_i \) for which each \( h_i \) is diagonal in the \( Z \) basis.
Then \( H \) is a classical Hamiltonian and \( \bB \)-\( \QkSAT \) is in \( \NP \).

We are interested in the complexity of basis-restricted \( \QkSAT \) when we allow for more than one ``type'' of basis.
In particular, is there some setting in which the problem already becomes \( \QMA \)-hard, even though the number of basis types is small?
We answer this question in the following sections by considering the problem \( \XZQkSAT \),
which we define as basis-restricted \( \QkSAT \) problem where
each \( h_i \) is diagonal in either the \( Z \) or the \( X \) basis.

\section{\( \XZQsixSAT \) is in \( \QMAone \)}

\begin{figure}[h]
  \caption{Decomposition of \( \Gamma(CCX) \) and \( \Gamma\HH \) into gates from \( \gs \).}
  \label{fig:control}
\centering
\includegraphics[width=0.21\textwidth]{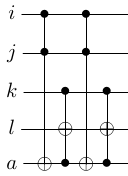}
\qquad
\includegraphics[width=0.5\textwidth]{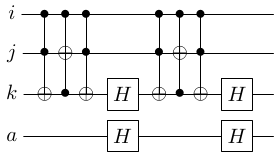}
\end{figure}

In this section, we begin the proof of \Cref{thm:main}
by showing that \( \XZQsixSAT \) is in \( \QMA^{\gs} \).
We start with a useful property of the gate set \( \gs \).
For gate \( U \), let \( \Gamma(U) \) denote the controlled-\( U \) gate.
\begin{lem}
  \label{lem:control}
  For every \( U \in \gs \), \( \Gamma(U) \) can be exactly implemented using gates in \( \gs \)
  (using an ancilla qubit, which can be in any state and will be left unchanged after the simulation).
\end{lem}
\begin{proof}
  \( \Gamma(X) \) and \( \Gamma(CX) \) are already included in \( \gs \).
  Let \( a \) denote an ancilla qubit.
  \( \Gamma(C C X) \) on control qubits \( i \),\( j \),\( k \) and target qubit \( l \) can be decomposed as
  \begin{align*}
    \Gamma(C C X)_{i,j,k,l} \otimes I_a = C C X_{i,j,a} C C X_{k,a,l} C C X_{i,j,a} C C X_{k,a,l}.
  \end{align*}
  
 For \( \Gamma\HH \) with control qubit \( i \) and target qubits \( j \),\( k \), we first give a decomposition using the \( \Gamma(\swap) \) gate.
  \begin{align*}
      \Gamma\HH_{i,j,k} \otimes I_a = \Gamma(\swap)_{i,j,k} \HH_{k, a} \Gamma(\swap)_{i,j,k} \HH_{k, a}.
  \end{align*}
  Then, since \( \Gamma(AB) = \Gamma(A)\Gamma(B) \) for any gates \( A \) and \( B \),
  the identity \( \swap_{i,j} = CX_{i,j} CX_{j,i} CX_{i,j} \) gives a decomposition of \( \Gamma(\swap) \) in terms of \( C C X \)'s.
  This completes the proof.
  We illustrate these constructions in Figure \ref{fig:control}.
\end{proof}

\begin{thm}
  \( \XZQsixSAT \) is in \( \QMAone^{\gs} \).
\end{thm}
\begin{proof}
  First, without loss of generality, we can build a \( \QMAone^{\gs} \) verification circuit which allows intermediate \( Z \) basis measurements and quantum operations conditioned on the measurement results.
  This is because a conditionally applied gate \( U \) can be replaced by the controlled operation \( \Gamma(U) \), removing the intermediate measurement.
  Lemma \ref{lem:control} then says that \( \Gamma(U) \) can then be decomposed back into the gate set \( \gG \).
  Thus, a circuit with intermediate measurements can be rewritten as a circuit with just a single measurement at the end of the computation, as in our definition of \( \QMAone \).
  Also note that classical processing can be performed freely, since our gate set includes \( C C X \), which is universal for classical computation.

  We use the verification procedure of Gosset and Nagaj \cite{GN16Quantum3SATQMA1Complete}, which is correct as long as the following condition holds:
  \begin{description}
  \item[(*)] There is an efficient quantum algorithm using gates in \( \gG \) which, given a projector \( \Pi \in \sS \), exactly performs the projective measurement \( \set{I - \Pi, \Pi} \) on a given state \( \ket{\psi} \).
  \end{description}
  Recall that \( \sS \) is the set of Hermitian \( 6 \)-local projectors which are diagonal in either the \( Z \) basis or the \( X \) basis.

  We briefly sketch how the \( \QMAone \) protocol works assuming (*) holds.
  First, choose a random projector term \( \Pi \) in the Hamiltonian \( H \) (random bits can be obtained by applying \( X \) to a \( \ket{0} \) ancilla then measuring).
  Then, perform the projective measurement in (*) on the proof state \( \ket{\psi} \).
  Accept when the outcome is \( I - \Pi \), and reject otherwise.
  The probability of accepting is \( 1 - \bra{\psi} \Pi \ket{\psi} \).
  When \( H \) has a zero-energy state, the protocol accepts this state with probability 1, showing perfect completeness.
  If instead \( \bra{\psi} H \ket{\psi} \ge 1\) for all \( \ket{\psi} \), the protocol will reject with probability at least \( 1 / \poly(n) \), which can be amplified to give the desired soundness; the full analysis is given in \cite{GN16Quantum3SATQMA1Complete}.

  We now prove that (*) holds.
  A projector \( \Pi \in \sS\) can be specified by a set of six qubits \( Q \) on which it acts nontrivially,
  a change of basis matrix \( V \) (which is either \( I \) or a tensor product of \( \had \)'s),
  and a set of supported strings \( S \subseteq \B^6 \), such that
  \begin{align*}
  \Pi = \sum_{z \in S} V \ktbr{z}{z} V.
  \end{align*}

  The algorithm first applies \( V \) on \( \ket{\psi} \), by applying \( \had \) on each of the specified qubits (using the \( \had \otimes \had \) gate, with the second \( \had \) acting on an otherwise unused ancilla qubit).
  It then measures the qubits in \( Q \) in the \( Z \) basis.
  Let \( z^\star \) be the measurement outcome.
  The algorithm returns outcome \( \Pi \) if \( z^\star \in S \), and otherwise returns outcome \( I - \Pi\).
  The probability of getting outcome \( \Pi \) is
  \begin{align*}
  \sum_{z \in S} \norm{\ktbr{z}{z} V \ket{\psi}}^2 = \bra{\psi} \Pi \ket{\psi},
  \end{align*}
  as desired. This completes the proof.
\end{proof}

\section{Circuit-to-XZ-Hamiltonian reduction}

In this section, we discuss some preliminaries for the \( \QMAone \)-hardness proof of \Cref{sec:hardness}.

\subsection{Two gate sets for \( \QMAone \)}
Recall that our chosen gate set for \( \QMAone \) is \( \gs = \set{X, CX, CCX, \had \otimes \had} \).
Define a new gate set
\[ \ngs = \set{X, CZ, CCZ, \hat{G}} ,\]
where \( \hat{G} \) is defined as the two-qubit operation equivalent to
\[ \hat{G} = \HH CZ \HH  .\]
We show that \( \ngs \) and \( \gs \) are interchangeable gate sets, in the following sense:
\begin{lem}
  \label{lem:ngs-equal}
  \( \QMAone^{\gs} = \QMAone^{\ngs} \).
\end{lem}
\begin{proof}
  We start by showing that \( \QMAone^{\ngs} \subseteq \QMAone^{\gs} \).
  It suffices to show that each gate in \( \ngs \) can be exactly written as a (constant length) sequence of gates from \( \gs \).
  Such a simulation can freely use the \( \QMAone \) verifier's ancilla register.
  In the sequences we construct, the ancilla qubits can be in any state and will be left unchanged after the gate simulation.
  Also note that \emph{exact} simulation is critical:
  merely approximating a gate using another gate set would not necessarily preserve the \( \QMAone \) protocol's perfect completeness.

To set some notation,
for a gate \( U \) and a set of gates \( S \), we say that \( U \in \overline{S} \) if \( U \) can be exactly written as a sequence of gates from \( S \).
The following closure property is immediate:
if \( S' \subseteq \overline{S}\) and \( U \in \overline{S'} \),
then \( U \in \overline{S} \).
In particular, if \( V \in \overline{S} \) and \( U \in \overline{S \cup \set{V}} \), then \( U \in \overline{S} \).

We now consider the gates in \( \ngs \).
The \( X \) gate is already in \( \gs \).
We can simulate the \( CZ \) gate on qubits \( i,j \) with an ancilla qubit \( a \),
using the circuit identity
\[ CZ_{i,j} \otimes I_a = \HH_{j,a} CX_{i,j} \HH_{j,a}.\]
This puts \( CZ \in \overline{\gs} \).
A similar identity
\[ CCZ_{i,j,k} \otimes I_a = \HH_{k,a} CCX_{i,j,k} \HH_{k,a} \]
puts \( CCZ \in \overline{\gs} \).
Finally, \( \hat{G} \in \overline{\gs \cup \set{CZ}} \) by definition,
so by the closure property we have \( \hat{G} \in \overline{\gs} \).

We now show that \( \QMAone^{\gs} \subseteq \QMAone^{\ngs} \) using the same approach.
We introduce a useful intermediate gate:
define \( \hat{F} \) to be the two-qubit operation equivalent to
\[ \hat{F} = \swap \HH .\]
We observe that \( \hat{F} = \hat{G} \cdot CZ \cdot \hat{G} \) using the circuit
\def\temp{0.07}
\[
  \centericg{\temp}{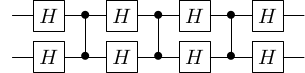}\enspace= \centericg{\temp}{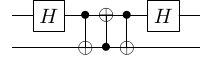} \enspace= \centericg{\temp}{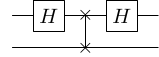} \enspace= \centericg{\temp}{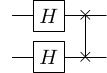},
\]
where we used that \( \had_j CZ_{i,j} \had_j = CX_{i,j} \) and \( \swap_{i,j} = CX_{i,j} CX_{j,i} CX_{i,j}  \).
This puts \( \hat{F} \in \overline{\ngs} \).
Next, using that \( \had \otimes \had \) and \( \swap \) commute,
we have that
\[ CX_{i,j} \otimes I_a = \hat{F}_{j,a} CZ_{i,a} \hat{F}_{j,a}, \]
since
\def\temp{0.1}
\[
  \centericg{\temp}{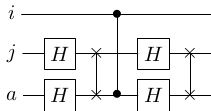} \enspace= \centericg{\temp}{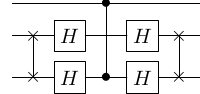} \enspace= \centericg{\temp}{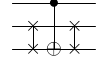} \enspace= \centericg{\temp}{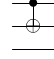}.
\]
This puts \( CX \in \overline{\ngs \cup \set{\hat{F}}} \), which implies that \( CX \in \overline{\ngs} \) by the closure property.
Similarly, the identity
\[ CCX_{i,j,k} \otimes I_a = \hat{F}_{k,a} CCZ_{i,j,a} \hat{F}_{k,a} \]
puts \( CCX \in \overline{\ngs} \).
Finally, note that \( \had \otimes \had = \swap \cdot \hat{F} \).
Writing \( \swap \) as three \( CX \)'s,
we have that \( \had \otimes \had \in \overline{\set{\hat{F}, CX}} \),
so \( \had \otimes \had \in \overline{\ngs} \) by the closure property.
\end{proof}

We find it useful to change to the gate set \( \ngs \) since each gate in \( \ngs \) has the following nice properties, which are straightforward to check:
\begin{lem}
  \label{lem:ngs-props}
  Each gate in \( \ngs \) is Hermitian, 3-local, and diagonal in either the \( Z \) basis or the \( X \) basis.
\end{lem}
These properties are used in the proof that \( \XZQsixSAT \) is \( \QMAone \)-hard, when we show that the Hamiltonian term \( H_{\mathrm{prop}} \) is indeed a valid instance of \( \XZQsixSAT \) (\Cref{lem:hprop-valid}).
\subsection{The Kitaev circuit-to-Hamiltonian construction}

We now sketch the original Kitaev construction and note which parts of the construction carry through to the proof of \Cref{thm:main} unchanged.

Let \( (L_{\mathrm{yes}}, L_{\mathrm{no}}) \). be a promise problem in \( \QMAone \),
and let \( V_x \) be the corresponding circuit which verifies a length \( n \) bitstring \( x \), given a proof state \( \ket{\psi} \).
Recall that the verification start from initial state
\( \ket{\mathrm{init}} = \ket{\psi} \ket{0 \cdots 0}_{\mathrm{anc}} \).
Let \( \hH \) denote the Hilbert space in which the initial state lives.
Let \( Q_{\mathrm{proof}} \) and \( Q_{\mathrm{anc}} \) denote the indices of the qubits in the proof and ancilla registers.
Write the circuit as \( V_x = U_T \cdots U_1 \), where each \( U_i \) is a gate from the chosen gate set for \( \QMAone \).

The Kitaev construction is an efficient mapping from the verification circuit \( V_{x} \) to a Hamiltonian
\[ H = H_{\mathrm{in}} + H_{\mathrm{out}} + H_{\mathrm{prop}} + H_{\mathrm{format}}. \]
\( H \) acts on Hilbert space \( \hH \otimes \hH_{\mathrm{clock}} \),
where \( \hH_{\mathrm{clock}} \) is an additional \( T \)-qubit \emph{clock register}.
The \emph{Kitaev clock states} \( \ket{\hat{t}} \in \hH_{\mathrm{clock}}, t \in \set{0, \dots, T} \) are defined as \( \ket{\hat{t}} = \ket{\clockfont{1}^t \clockfont{0}^{T-t}} \) (note that the definition of clock states will differ in our construction).
This reduction has the following properties:

\begin{description}
\item[Completeness] If \( x \in L_{\mathrm{yes}} \), then the \emph{history state}
  \begin{align}
  \label{eqn:hist}
    \ket{\mathrm{hist}} = \frac{1}{\sqrt{T+1}}\sum_{t = 0}^T U_t \cdots U_1 \ket{\mathrm{init}} \otimes \ket{\hat{t}}
  \end{align}
  is a zero-energy state of \( H \).
\item[Soundness] If \( x \in L_{\mathrm{no}} \), then any state \( \ket{\varphi} \) has at least \( \frac{1}{\poly(n)} \) energy with respect to \( H \).
\item[Instance of \( \QSAT \)] \( H \) is a sum of projectors.
  Moreover, if each gate in the chosen gate is \( k \)-local,
  then \( H \) is \( (k+3) \)-local.
\end{description}

Each term in the Hamiltonian represents a constraint on the proof state (which is claimed to be the history state);
an energy penalty is given if the constraint is not met.

\subsection{\( H_{\mathrm{prop}} \) and \( H_{\mathrm{format}} \)}
\label{sec:hprop}

Our construction differs from the Kitaev construction in the definition of \( H_{\mathrm{prop}} \) and \( H_{\mathrm{format}} \), which we define in this section.
First, we make a simple observation that the verification circuit \( V_x \) can be assumed to be in a standard form.

\begin{lem}
  \label{lem:std-form}
  Without loss of generality, \( V_x = U_T \cdots U_1 \) (where \( U_i \in \ngs \)) satisfies the following properties: 
  \begin{itemize}
  \item \( T \) is odd.
  \item \( U_i \) is diagonal in the \( X \) basis when \( i \) is odd.
  \item \( U_i \) is diagonal in the \( Z \) basis when \( i \) is even.
  \end{itemize}
\end{lem}
\begin{proof}
  Let \( W_{\tau} \cdots W_1 \) be any \( \QMAone \) verification circuit with gates \( W_j \) from the gate set \( \ngs \).
  Recall that by \Cref{lem:ngs-props}, each \( W_j \) is diagonal in either the \( X \) basis or the \( Z \) basis.
  By padding the circuit with identity operations,
  we can construct an equivalent circuit \( U_T \cdots U_1 \) of \( T = 2 \tau + 1 \) gates which has the desired form.
  Concretely, for each \( j \in [\tau] \), if \( W_j \) is diagonal in the \( X \) basis,
  set \( U_{2j - 1} = W_j \), and otherwise set \( U_{2j} = W_j \).
  The gates \( U_i \) which remain undefined by this procedure are set to be the identity operation.
\end{proof}

We now define the clock states in our construction,
which differ from those for the original Kitaev Hamiltonian.
Let
\begin{align*}
  \ket{\hat{0}} &= \ket{\clockfont{0+0+0+}\tcd\clockfont{0}} \\
  \ket{\hat{1}} &= \ket{\clockfont{1+0+0+}\tcd\clockfont{0}} \\
  \ket{\hat{2}} &= \ket{\clockfont{1-0+0+}\tcd\clockfont{0}} \\
  \ket{\hat{3}} &= \ket{\clockfont{1-1+0+}\tcd\clockfont{0}} \\
  \ket{\hat{4}} &= \ket{\clockfont{1-1-0+}\tcd\clockfont{0}} \\
                &\,\,\,\vdots\\
  \ket{\hat{T}} &= \ket{\clockfont{1-1-1-}\ttcd\clockfont{1}}.
\end{align*}
To remove ambiguity, we call these states \emph{good clock states},
and refer to the clock states of the Kitaev Hamiltonian as \emph{Kitaev clock states}.
Unlike a Kitaev clock state, which has \( Z \) basis states on all qubits,
a good clock state consists of alternating \( Z \) and \( X \) basis states.
On the \( i \)-th ``tick'' of the clock,
the \( \lket{0} \) in the \( i \)-th position changes to a \( \lket{1} \) if \( i \) is odd
and changes from \( \lket{+} \) to \( \lket{-} \) if \( i \) is even.

Our new propagation term has the form \( H_{\mathrm{prop}} = \sum_{t=1}^{T} H_{\mathrm{prop}, t} \), with
\begin{align*}
  H_{\mathrm{prop},t}=
  \begin{cases}
    \frac{1}{2} [I_{\hH} \otimes ( \lktbr{0+}{0+}_{1,2} + \lktbr{1+}{1+}_{1,2} )\\
    \quad- U_1 \otimes \lktbr{1+}{0+}_{1,2} - U_1^\dagger \otimes \lktbr{0+}{1+}] &\text{if } t = 1, \\
    \frac{1}{2} [I_{\hH} \otimes ( \lktbr{1+0}{1+0}_{t-1,t,t+1} + \lktbr{1-0}{1-0}_{t-1,t,t+1} )\\
    \quad- U_t \otimes \lktbr{1-0}{1+0}_{t-1,t,t+1} - U_t^\dagger \otimes \lktbr{1+0}{1-0}_{t-1,t,t+1}] &\text{if } t \text{ is even}, \\
    \frac{1}{2} [I_{\hH} \otimes ( \lktbr{-0+}{-0+}_{t-1,t,t+1} + \lktbr{-1+}{-1+}_{t-1,t,t+1} )\\
    \quad- U_t \otimes \lktbr{-1+}{-0+}_{t-1,t,t+1} - U_t^\dagger \otimes \lktbr{-0+}{-1+}] &\text{if } t \text{ is odd, } t \ne 1, t \ne T, \\
    \frac{1}{2} [I_{\hH} \otimes ( \lktbr{-0}{-0}_{T-1,T} + \lktbr{-1}{-1}_{T-1,T} ) \\
    \quad- U_T \otimes \lktbr{-1}{-0}_{T-1,T} - U_T^\dagger \otimes \lktbr{-0}{-1}_{T-1,T}] &\text{if } t = T.
  \end{cases}
\end{align*}
Importantly, when the gates \( U_t \) come from \( \ngs \), \( H_{\mathrm{prop}, t} \) satisfies the conditions imposed on instances of \( \XZQsixSAT \).
\begin{lem}
  \label{lem:hprop-valid}
  For gates \( U_t \in \ngs  \), \( H_{\mathrm{prop}, t} \) is a \( 6 \)-local projector which is diagonal in either the \( Z \) basis or the \( X \) basis.
\end{lem}
\begin{proof}
  It is straightforward to check that \( H_{\mathrm{prop}, t} \) is a projector, i.e. \( H_{\mathrm{prop},t}^2 = H_{\mathrm{prop},t} \).
  First, consider when \( t \) is odd and \( t\ne1, T \).
  By \Cref{lem:ngs-props}, \( U_t = U_t^\dagger \), so by factoring out qubits \( t-1 \) and \( t+1 \) in the clock register, we have
  \begin{align*}
    H_{\mathrm{prop}, t} &= \lktbr{-+}{-+}_{t-1, t+1} \otimes \frac{1}{2} \left[ I_{\hH} \otimes \left( \lktbr{0}{0}_t + \lktbr{1}{1}_t \right) - U_t \otimes \left( \lktbr{1}{0}_t + \lktbr{0}{1}_t  \right)   \right]  \\
                       &= \lktbr{-+}{-+}_{t-1, t+1} \otimes \frac{1}{2} \left[I_{\hH} \otimes I_t - U_t \otimes X_t \right],
  \end{align*}
  where \( I_t \) and \( X_t \) are the identity and Pauli \( X \) operators on clock qubit \( t \).
  By \Cref{lem:std-form}, \( U_t \) is diagonal in the \( X \) basis, since \( t \) is odd.
  In this form, it is clear what basis diagonalizes \( H_{\mathrm{prop}, t} \): the change-of-basis matrix applies \( \had \) on all qubits.
  Thus, \( H_{\mathrm{prop}, t} \) is diagonal in the \( X \) basis.
  Moreorer, by \Cref{lem:ngs-props}, \( U_t \) is \( 3 \)-local, so \( H_{\mathrm{prop}, t} \) is \( 6 \)-local.

  By a similar argument, \( H_{\mathrm{prop},t} \) is diagonal in the \( Z \) basis for odd \( t \) (including \( t=1 \) and \( t=T \)),
  noting that \( \lktbr{-}{+}_t + \lktbr{-}{+}_t = Z_t \).
\end{proof}

We now define the formatting Hamiltonian term \( H_{\mathrm{format}} = H_{\mathrm{format},X} + H_{\mathrm{format},Z} \), where
\begin{align*}
  H_{\mathrm{format}, X} &= I_{\hH} \otimes \left( \lktbr{+-}{+-}_{2,4} + \lktbr{+-}{+-}_{4,6} + \cdots + \lktbr{+-}{+-}_{T-3, T-1} \right),\\
  % \sum_{i=1}^{\frac{T-3}{2}} \lktbr{+-}{+-}_{2i, 2i + 2} =
  H_{\mathrm{format}, Z} &= I_{\hH} \otimes \left( \lktbr{01}{01}_{1,3} + \lktbr{01}{01}_{3,5} + \cdots + \lktbr{01}{01}_{T-2, T} \right).
  % \sum_{i=0}^{\frac{T-3}{2}} \lktbr{01}{01}_{2i+1, 2i+3} =
\end{align*}
The \( Z \) format terms give an energy penalty unless the odd clock qubits form a Kitaev clock state.
Likewise, the \( X \) format terms only allow Kitaev clock states on the even clock qubits (under the relabeling \( \lket{0} \to \lket{+} \) and \( \lket{1} \to\lket{-} \)).

It is clear that every good clock state is a zero-energy state of \( H_{\mathrm{format}}\). However, note that that there are some zero-energy states of \( H_{\mathrm{format}} \) which are not good clock states!
We refer to these states as \emph{fake clock states}. Intuitively, these are states where the \( X \) and \( Z \) clocks are each valid Kitaev clock states, but they are not properly synchronized with each other.
In the following section, we will show the ground space of \( H \) has no support on fake clock states,
even though they are zero-energy states of \( H_{\mathrm{format}} \).
We also use the term \emph{bad clock states} to refer to non-zero energy states of \( H_{\mathrm{format}} \).

\subsection{\( H_{\mathrm{in}} \) and \( H_{\mathrm{out}} \)}

The terms \( H_{\mathrm{in}} \) and \( H_{\mathrm{out}} \) of the Kitaev construction remain unchanged in our construction, and we state them here:
\begin{align}
  \label{eq:hin}
  H_{\mathrm{in}} &= \sum_{i \in Q_{\mathrm{anc}}} \ktbr{1}{1}_i \otimes \lktbr{0}{0}_1,  \\
  \label{eq{hout}}
  H_{\mathrm{out}} &= \ktbr{0}{0}_1 \otimes \lktbr{1}{1}_T.
\end{align}

Within the subspace of clock states, the local check \( \lktbr{0}{0}_1 \) projects onto the subspace spanned by \( \ket{\hat{0}} \).
Thus, \( H_{\mathrm{in}} \) assigns an energy penalty to states whose \( \ket{\hat{0}} \) clock component contains ancilla qubits which are not initialized to \( \ket{0} \).
Likewise, \( H_{\mathrm{out}} \) assigns an energy penalty when the \( \ket{\hat{T}} \) clock component of the state does not have a \( \ket{1} \) on the first qubit of \( \hH \) (recall that this is qubit measured at the end of the \( \QMAone \) verification).
Note that \(\bra{\mathrm{hist}} H_{\mathrm{in}} \ket{\mathrm{hist}} = \bra{\mathrm{hist}} H_{\mathrm{out}} \ket{\mathrm{hist}} = 0\).

\section{\( \XZQsixSAT \) is \( \QMAone \)-hard}
\label{sec:hardness}
In this section, we show that \( \XZQsixSAT \) is \( \QMAone^{\ngs} \)-hard, which suffices to prove \Cref{thm:main} by \Cref{lem:ngs-equal}.

Completeness is immediate: it is straightforward to check that in the ``yes'' case, the history state given by \Cref{eqn:hist} (where \( \ket{\hat{t}} \) now refers to a good clock state, not a Kitaev clock state, and \( \ket{\psi} \) is the accepting proof of the \( \QMAone \) verifier) is a zero-energy state of our Hamiltonian \( H \).

In the remainder of this section,
we prove soundness, i.e. that in the ``no'' case,
any state in \( \hH \otimes \hH_{\mathrm{clock}} \) has energy at least \( 1 / \poly(n) \) with respect to \( H \).
In other words, the smallest eigenvalue \( \lambda_{\min}(H) \) is at least \( 1 / \poly(n) \).
Recall that \( T \) is odd by \Cref{lem:std-form},
and let \( T = 2 \tau + 1 \).

\paragraph{Setup and notation.}
We first partition \( \hH \otimes \hH_{\mathrm{clock}} \) into subspaces corresponding to good, fake, and bad clock states.
For strings \( a = a_1 \cdots a_{j+1} \) and \( b = b_1 \cdots b_j \), we
use \( \itl{a}{b} = a_1b_1a_2b_2\cdots a_jb_j a_{j+1}\) to denote the string formed by interleaving \( a \) and \( b \).
Define the set of strings \( S = \set{\itl{a}{b} : a \in \set{\clockfont{0}, \clockfont{1}}^{\tau+1}, b \in \set{\clockfont{+}, \clockfont{-}}^{\tau}} \).
The states \( \set{\ket{s} : s \in S} \) form a basis for \( \hH_{\mathrm{clock}} \).
We partition \( S \) into subsets \( S_{\mathrm{good}} \), \( S_{\mathrm{fake}} \), and \( S_{\mathrm{bad}} \).
First, define
\[ S_{\mathrm{format}} = \set{\itl{\clockfont{1}^{t_Z}\clockfont{0}^{\tau + 1 - t_Z}}{\clockfont{-}^{t_X}\clockfont{+}^{\tau - t_X}} : t_Z \in \set{0, \dots, \tau + 1}, t_X \in \set{0, \dots, \tau}}. \]
Let \( S_{\mathrm{good}} \subseteq S_{\mathrm{format}} \) contain the strings of the above form
which satisfy either \( t_Z = t_X \) or \( t_Z = t_X + 1 \).
Let \( S_{\mathrm{fake}} = S_{\mathrm{format}} \setminus S_{\mathrm{good}} \) and \( S_{\mathrm{bad}} = S \setminus S_{\mathrm{format}} \).
It is clear that \( S_{\mathrm{good}} \) and \( S_{\mathrm{fake}} \) correspond to good and fake clock states respectively,
while \( S_{\mathrm{bad}} \) corresponds to non-zero-energy states of \( H_{\mathrm{format}} \).

Let \( \hH_{\mathrm{good}} \) be the subspace of \( \hH_{\mathrm{clock}} \) spanned by \( \set{\ket{s} : s \in S_{\mathrm{good}}} = \set{\ket{\hat{0}}, \ket{\hat{1}}, \dots, \ket{\hat{T}}} \).
Note that \( \hH_{\mathrm{good}}^\perp \) (i.e. the orthogonal complement of \( \hH_{\mathrm{good}} \) in \( \hH_{\mathrm{clock}} \)) is the span of \( \set{\ket{s} : s \in S_{\mathrm{fake}} \cup S_{\mathrm{bad}}} \).

\paragraph{Factoring out the good subspace.}
We first show that \( \hH \otimes \hH_{\mathrm{good}} \) is an invariant subspace of \( H \).
This is clear for \( H_{\mathrm{in}} \), \( H_{\mathrm{out}} \), and \( H_{\mathrm{format}} \) since each term acts as identity or zero on \( \hH_{\mathrm{clock}} \).
For \( H_{\mathrm{prop}} \), we see that for \( t \in [T] \) and \( \ket{\varphi} \in \hH \),
\begin{align}
  \label{eqn:hprop}
  H_{\mathrm{prop},t} \ket{\varphi} \ket{\widehat{t-1}} &= \ket{\varphi} \ket{\widehat{t-1}} + U_t\ket{\varphi} \ket{\hat{t}}, \\
  H_{\mathrm{prop},t} \ket{\varphi} \ket{\hat{t}} &= U_t^\dagger \ket{\varphi} \ket{\widehat{t-1}} + \ket{\varphi} \ket{\hat{t}}, \\
  H_{\mathrm{prop},t} \ket{\varphi} \ket{\hat{k}} &= 0 \text{ for } k \in \set{0, \dots, T} \setminus \set{t-1, t},
\end{align}
as desired.

We can now write
\begin{align}
\label{eq:ortho-decomp}
\hH \otimes \hH_{\mathrm{clock}} = \hH \otimes \left( \hH_{\mathrm{good}} \oplus \hH_{\mathrm{good}}^\perp \right) = \left( \hH \otimes \hH_{\mathrm{good}} \right) \oplus \left( \hH \otimes \hH_{\mathrm{good}}^\perp \right).
\end{align}
In fact, \( \hH \otimes \hH_{\mathrm{good}} \) and \( \hH \otimes \hH_{\mathrm{good}}^\perp \) are orthogonal,
by the orthogonality of \( \hH_{\mathrm{good}} \) and \( \hH_{\mathrm{good}}^\perp \).
It follows from \Cref{eq:ortho-decomp} that \( \hH \otimes \hH_{\mathrm{good}}^\perp = (\hH \otimes \hH_{\mathrm{good}})^\perp \) (the orthogonal complement of \( \hH \otimes \hH_{\mathrm{good}} \) in \( \hH \otimes \hH_{\mathrm{clock}} \)).

We now apply the following linear algebra fact:
\begin{lem}
  \label{lem:orthocomp-invar}
  If \( W  \) be a \( O \)-invariant subspace of Hilbert space \( V \),
  then \( W^\perp \) is \( O^\dagger \)-invariant.
\end{lem}
\begin{proof}
  Let \( z \in W^\perp \).
  Denoting the inner product by \( (\cdot,\cdot) \),
  we have for any \( w \in W \) that \( (w, O^\dagger z) = (Ow, z) = 0 \),
  since \( Ow \in W \) and \( z \in W^\perp \).
  Thus \( O^\dagger z \in W^\perp \), as desired.
\end{proof}
We conclude that \( \hH \otimes \hH_{\mathrm{good}}^\perp \) is \( H \)-invariant as well,
using \Cref{lem:orthocomp-invar} and the fact that \( H \) is Hermitian.
\( H \) can then be block-diagonalized, simplifying our analysis.
We get the eigenvalues of \( H \restriction_{\hH \otimes \hH_{\mathrm{clock}}}\) (i.e. \( H \) as an operator on \( \hH \otimes \hH_{\mathrm{clock}} \)) by taking the union of the eigenvalues of \( H \restriction_{\hH \otimes \hH_{\mathrm{good}}} \) and those of \( H \restriction_{\hH \otimes \hH_{\mathrm{good}}^\perp} \).
In particular,
\begin{equation} \lambda_{\min}(H \restriction_{\hH \otimes \hH_{\mathrm{clock}}}) = \min(\lambda_{\min}(H \restriction_{\hH \otimes \hH_{\mathrm{good}}}), \lambda_{\min}(H\restriction_{\hH \otimes \hH_{\mathrm{good}}^\perp})). \label{eq:lambdamin-across-sectors}
\end{equation}

We first lower bound \( \lambda_{\min}(H \restriction_{\hH \otimes \hH_{\mathrm{good}}}) \). Observe that \( H_{\mathrm{format}} \) has zero energy on good states, and the remaining part \( H_{\mathrm{in}} + H_{\mathrm{out}} + H_{\mathrm{prop}} \) is identical to the Hamiltonian in Kitaev's original reduction \cite{KSV02ClassicalQuantumComputation} after doing a change of basis which applies \( \had \) on all even clock qubits.
Since the eigenvalues are invariant under change of basis,
the lower bound from the analysis in \cite{KSV02ClassicalQuantumComputation} carries over directly:
we have that \begin{equation} 
\lambda_{\min}(H \restriction_{\hH \otimes \hH_{\mathrm{good}}}) \ge 1 / \poly(n). \label{eq:lambdamin-good} \end{equation}
In the next part of this section, we find a lower bound when \( H \) acts on \( \hH \otimes \hH_{\mathrm{good}}^\perp \).

\paragraph{Bounding the energy outside the good subspace.}
First, note that for any \( \ket{\varphi} \in \hH \otimes \hH_{\mathrm{good}}^\perp \),
\[ \bra{\varphi} (H_{\mathrm{in}} + H_{\mathrm{out}} + H_{\mathrm{prop}} + H_{\mathrm{format}}) \ket{\varphi} \ge \bra{\varphi} (H_{\mathrm{prop}} + H_{\mathrm{format}}) \ket{\varphi}, \]
since \( H_{\mathrm{in}} + H_{\mathrm{out}} \) is positive definite.
It thus suffices to argue that \( H_{\mathrm{prop}} + H_{\mathrm{format}} \) has energy at least \( 1 / \poly(n) \) on \( \hH \otimes \hH_{\mathrm{good}}^\perp \)
(in fact, we will show that the energy is \( \Omega(1) \)).

\newcommand{\fake}{\mathrm{fake}}
\newcommand{\bad}{\mathrm{bad}}
\newcommand{\good}{\mathrm{good}}
\newcommand{\format}{\mathrm{format}}
\newcommand{\prp}{\mathrm{prop}}

Let us define $\hH_{\fake}$ and $\hH_{\bad}$ in terms of $S_\fake$ and $S_\bad$ analogously to  $\hH_{\good}$, so that $\hH_{\good}^\perp = \hH_{\fake} \oplus \hH_{\bad}$. Suppose we have a state $\ket{\psi}$ supported only on $\hH_{\good}^\perp$. Let us write this state as  
\begin{equation}
    \ket{\psi} = \sum_{i} \ket{\psi_i} \otimes \ket{i}_{\mathrm{clock}},
\end{equation}
where the vectors $\ket{i}$ are clock basis states, and the vectors $\ket{\psi_i}$ are subnormalized. We are now going to bound $H_{\prp} + H_\format$ by decomposing it in terms of the components $\ket{\psi_i}$. Firstly, $H_{\format}$ is simple to bound:
\begin{equation}
    \bra{\psi} H_\format \ket{\psi} \geq \sum_{i \in S_\bad}  \| \ket{\psi_i} \|^2.
\end{equation}
This follows because $H_\format$ assigns an energy penalty of at least $1$ to every bad string. Next, let us look at a single propagation term $H_{\prp, t}$. We will assume that \( t \) is odd and $1 < t < T$ for simplicity, and also use the fact that all of our gates are self-adjoint to simplify expressions. We will write this propagation term as a sum of terms that look like graph Laplacians over the graph whose vertices consist of clock basis strings, and whose edges correspond to pairs of strings that are paired by $H_{\prp,t}$.

\begin{align}
   \bra{\psi} H_{\prp,t} \ket{\psi} &= \bra{\psi} \left[ \frac{1}{2} (I_\hH \otimes I_{t} - (U_t)_\hH \otimes X_t ) \otimes \lktbr{-+}{-+}_{t-1, t+1} \otimes I_{\mathrm{rest of clock}} \right]  \ket{\psi} \\
   &= \frac{1}{2} \sum_{i,j} \bra{\psi_i} \otimes \bra{i}_{\mathrm{clock}}\Big[(I_\hH \otimes I_t - (U_t)_\hH \otimes X_t) \otimes \lktbr{-+}{-+}_{t-1, t+1} \nonumber \\ 
   &\qquad\qquad\qquad\qquad\otimes I_{\mathrm{rest of clock}}\Big] \ket{\psi_j} \otimes \ket{j}_{\text{clock}} \\\
   &= \frac{1}{2} \sum_{(i,j) \in E_t} ( \ipr{\psi_i}{\psi_i} + \ipr{\psi_j}{\psi_j} - \bra{\psi_i} U_t \ket{\psi_j}  - \bra{\psi_j} U_t \ket{\psi_i}) \\
   &= \frac{1}{2} \sum_{(i,j) \in E_t} \| \ket{\psi_i} - U_t \ket{\psi_j} \|^2, 
   \end{align}
   where $E_t$ consists of all unordered pairs $(i,j)$ of strings in $S_\fake \cup S_\bad$ such that $i$ and $j$ agree at all positions except position $t$, \emph{disagree} at position $t$, and both strings have $\clockfont{-}$ in position $t-1$ and $\clockfont{+}$ in position $t+1$. (For other values of $t$ other than the ones we have considered here, $E_t$ may be defined analogously: we give a full definition encapsulating all edge cases in \Cref{def:hprop-graph}.) To simplify notation for what will follow, let us replace $U_t$ in the last line with $U_{ij}$: this is well-defined because it is easy to see that the sets $E_t$ are disjoint for different values of $t$. Thus, we have
   \begin{equation} \bra{\psi} H_{\prp, t} \ket{\psi} =\frac{1}{2} \sum_{(i,j) \in E_t} \| \ket{\psi_i} - U_{ij} \ket{\psi_j} \|^2. \end{equation}

   Putting these together, the total energy of $H_{\prp} + H_{\format}$ is
   \begin{equation}
   \bra{\psi} (H_{\prp} + H_{\format}) \ket{\psi}  \geq \sum_{(i,j) \in E} \frac{1}{2} \| \ket{\psi_i} - U_{ij} \ket{\psi_j} \|^2 + \sum_{i \in S_{\bad}} \|\ket{\psi_i} \|^2, \label{eq:energy-big-edge-sum}
    \end{equation}
    where $E = \bigcup_{t} E_t$ is the union of all the edges associated with all the terms  of $H_{\prp}$.

    To lower-bound this quantity, we need to show that the fake strings are well-connected to the bad strings, thus forcing a large amount of weight onto the second summation. To do this, we will use two facts shown in the next section to study the connected component structure of the graph $G$ whose vertices are $S_\fake \cup S_\bad$ and whose edges are $E$. By \Cref{lem:comp-fb}, every fake string is connected to at least one bad string. This means that the connected components of $G$ consist either exclusively bad strings, or a mixture of fake and bad strings: there are no components of $G$ containing only fake strings. Furthermore, by \Cref{lem:comp-onef}, each connected component contains at most one fake string. Thus, the components either consist entirely of bad strings, or exactly one fake string and at least one bad string. 
    
    Moreover, we can freely drop edges from the summation in \Cref{eq:energy-big-edge-sum}, and only lower the energy. We will try dropping all the ``bad to bad'' edges, so that the only remaining edges are between fake and bad strings. Let us introduce the notation $N(i)$ to refer to the set of all neighbors in the graph $G$ of a string $i$. Observe that by the facts about the component structure of $G$ mentioned in the previous paragraph, it holds that for distinct $i,j \in S_\fake$, the sets $N(i), N(j)$ are disjoint, so every string $x \in S_{\bad}$ is contained in at most one $N(i)$ for $i \in S_\fake$. Using this characterization, we write \Cref{eq:energy-big-edge-sum} after the bad-to-bad edges have been omitted as follows:
        \begin{align} \bra{\psi}(H_{\prp,t} + H_\format) \ket{\psi} \geq \sum_{i \in S_{\fake}} \left[ \sum_{j \in N(i)} \left(\frac{1}{2} \|\ket{\psi_i} - U_{ij} \ket{\psi_j}\|^2 + \|\ket{\psi_j}\|^2\right)\right] + \sum_{i \in S_{\mathrm{bad-leftover}}} \|\ket{\psi_i}\|^2, \label{eq:energy-components}
        \end{align}
    where $S_{\mathrm{bad-leftover}} = S_{\bad} \setminus \bigcup_{i \in S_{\fake}} N(i)$. 

    Now, because the neighborhoods $N(i)$ are disjoint for distinct $i \in S_\fake$, we see that the entire RHS of \Cref{eq:energy-components} can be written as a sum 
    \[ \sum_{i \in S_{\fake} \cup S_{\mathrm{bad-leftover}}} \bra{\psi} M_i \ket{\psi},\]
    where the matrices $M_i$ are Hermitian matrices that act on different factors of the clock space, so that
    \begin{equation}
        \bra{\psi} M_i \ket{\psi} = \begin{cases} 
        \sum_{j \in N(i)} (\frac{1}{2} \|\ket{\psi_i} - U_{ij} \ket{\psi_j}\|^2 + \|\ket{\psi_j}\|^2) & \text{if $i \in S_\fake$}, \\
        \|\ket{\psi_i} \|^2 & \text{if $i \in S_{\mathrm{bad-leftover}}$}.
        \end{cases}
    \end{equation}
    From this, it is easy to see that the minimum value of the RHS of \Cref{eq:energy-components} is simply the minimum eigenvalue of the matrices $M_i$, restricted to their associated factors. For $i \in S_{\mathrm{bad-leftover}}$, $M_i$ acts as the identity matrix on its corresponding clock sector, so its minimum eigenvalue is $1$. It thus remains to lower bound the minimum eigenvalue of $M_i$ when $i \in S_{\fake}$. By \Cref{cor:opt-by-hand-with-u}, this is at least $1/4$. So overall, we conclude that for $\ket{\psi}$ supported on $\hH_{\good}^\perp$, 
    \[\bra{\psi} (H_\prp + H_\format) \ket{\psi} \geq 1/4, \]
    and hence 
    \begin{equation} \lambda_{\min}(H \restriction_{\hH \otimes \hH_{\mathrm{good}}^\perp}) \geq 1/4. \label{eq:lambdamin-fakebad}
    \end{equation}
    Combining \Cref{eq:lambdamin-across-sectors} with \Cref{eq:lambdamin-good} and \Cref{eq:lambdamin-fakebad}, we conclude that 
    \[ \lambda_{\min}(H) \geq 1/\poly(n)\]
    as desired, establishing the soundness property.
\section{Soundness: technical lemmas}

In this section we prove the technical lemmas that were used in the soundness proof in the previous section.
\paragraph{The structure of \( H_{\mathrm{prop}} \)}
 We start with some useful facts about how strings in \( S_{\mathrm{fake}} \cup S_{\mathrm{bad}} \) are coupled together by \( H_{\mathrm{prop}} \). Our first step is to more formally define the graph $G$ capturing this coupling, which was introduced in the previous section.
\begin{defn}\label{def:hprop-graph}
  Define a graph \( G \) with vertex set \(S_{\mathrm{fake}} \cup S_{\mathrm{bad}} \) and edge set \( E = \cup_{i=1}^T E_i \), where
  \begin{itemize}
  \item \( (u,v) \in E_1 \) iff \( u_1 \ne v_1 \), \( u_2 = v_2 = \clockfont{+} \), and \( u_i = v_i \) for all \( i \ge 3 \).
  \item For \( t \in \set{2, 4, \cdots, T-3, T-1} \),
    \( (u,v) \in E_t \) iff \( u_t \ne v_t \), \( u_{t-1} = v_{t-1} = \clockfont{1} \), \( u_{t+1} = v_{t+1} = \clockfont{0} \),
    and \( u_i = v_i \) for all \( i \ne t-1,t,t+1 \).
  \item For \( t \in \set{3, 5, \dots, T-4, T-2} \),
    \( (u,v) \in E_t \) iff \( u_t \ne v_t \), \( u_{t-1} = v_{t-1} = \clockfont{-} \), \( u_{t+1} = v_{t+1} = \clockfont{+} \),
    and \( u_i = v_i \) for al \( i \ne t-1,t,t+1 \).
  \item \( (u,v) \in E_T \) iff \( u_T \ne v_T \), \( u_{T-1} = v_{T-1} = \clockfont{-} \), and \( u_i = v_i \) for all \( i \le T-2 \).
  \end{itemize}
For vertex \( v \) in \( G \),
let \( \cC(v) \) be the connected component of \( G \) which contains \( v \).
\end{defn}
It is clear by the definition of \( H_{\mathrm{prop}} \) that edges in \( G \) correspond to states which are connected under the action of \( H_{\mathrm{prop}} \) (in the sense of \Cref{eqn:hprop}).
We now prove some important lemmas about the connectivity of \( G \) (\Cref{lem:comp-fb}, \Cref{lem:comp-onef}), which allow us to understand \( H_{\mathrm{prop}} \).

\begin{lem}
  \label{lem:comp-fb}
  For every \( f \in S_{\mathrm{fake}} \),
  there is some \( b \in S_{\mathrm{bad}} \) such that \( (f,b) \in E \).
\end{lem}
\begin{proof}
  Let \( f \in S_{\mathrm{fake}} \).
  Then \( f = \itl{\clockfont{1}^{t_Z}\clockfont{0}^{\tau + 1 - t_Z}}{\clockfont{-}^{t_X}\clockfont{+}^{\tau - t_X}} \) for some \( t_Z \in \set{0, \dots, \tau + 1} \), \( t_X \in \set{0, \dots, \tau} \)
  satisfying \( t_Z \ne t_X \) and \( t_Z \ne t_X + 1 \).
  We case on \( t_X \).
  First, suppose \( t_X = 0\).
  Then \( t_Z \ge 2\), so \( f \) starts with \( \clockfont{1+1} \).
  Define \( b \) to be the same string as \( f \),
  but with the first position changed to a \( \clockfont{0} \).
  Then \( (f, b) \in E_1 \),
  by the definition of \( E_1 \).
  Note that there is a \( \clockfont{0} \) followed by a \( \clockfont{1} \) on the odd terms of \( b \),
  which puts \( b \in S_{\mathrm{bad}} \), as desired.

  Suppose instead that \( t_X \in [\tau-1] \).
  We can subdivide into two cases:
  \( t_Z \) satisfies either \( t_Z \le t_X - 1 \) or \( t_Z \ge t_X + 2 \).
  If \( t_Z \le t_X - 1 \),
  then \( f \) looks like
  \begin{align*}
    &\spc\spc\spc\spc\wild\spc\wild\spc\clockfont{0}\spc\clockfont{0}\spc\clockfont{0}\spc\clockfont{0}\\[-13pt]
    &\tcd\spc\spc\spc\spc\spc\spc\spc\spc\spc\spc\spc\tcd,\\[-13pt]
    &\spc\spc\spc\spc\spc\clockfont{-}\spc\clockfont{-}\spc\underline{\clockfont{-}}\spc\clockfont{+}\spc\clockfont{+}\spc
  \end{align*}
  where we have have underlined position \( 2 t_X \) and offset the odd and even terms for clarity,
  with \( \wild \) denoting either a \( \clockfont{0} \) or a \( \clockfont{1} \).
  Let \( b \) be the same as \( f \) but with position \( 2t_X + 1 \) changed to a \( \clockfont{1} \).
  Then \( (f, b) \in E_{2 t_X + 1} \),
  and \( b \in S_{\mathrm{bad}} \) since \( b_{2 t_X -1} = \clockfont{0} \) while \( b_{2 t_X + 1} = \clockfont{1} \).
  
  If instead \( t_Z \ge t_X + 2 \), then \( f \) looks like
  \begin{align*}
    &\spc\spc\spc\spc\clockfont{1}\spc\clockfont{1}\spc\clockfont{1}\spc\clockfont{1}\spc\wild\spc\wild\\[-13pt]
    &\tcd\spc\spc\spc\spc\spc\spc\spc\spc\spc\spc\spc\tcd,\\[-13pt]
    &\spc\spc\spc\spc\spc\clockfont{-}\spc\underline{\clockfont{-}}\spc\clockfont{+}\spc\clockfont{+}\spc\clockfont{+}\spc
  \end{align*}
  where we have underlined position \( 2 t_X \).
  Define \( b \) to be \( f \) but with position \( 2 t_X + 1 \) changed to a \( \clockfont{0} \).
  Once again, we have \( (f,b) \in E_{2 t_X + 1} \),
  and \( b \in S_{\mathrm{bad}} \) since \( b_{2 t_X + 1} = \clockfont{0} \) while \( b_{2 t_X + 3} = \clockfont{1} \).
  
  Finally, consider when \( t_X = \tau \).
  Then \( t_Z \le \tau - 1 \),
  so \( f \) ends in \( \clockfont{0-0} \).
  Let \( b \) be the same as \( f \) but with the final position changed to a \( \clockfont{1} \).
  Then \( (f, b) \in E_{T} \) and \( b \in S_{\mathrm{bad}} \),
  completing the proof.
\end{proof}

We now prove an intermediate ``locking'' lemma.

\begin{lem}
  \label{lem:lock}
  Let \( s \in S_{\mathrm{fake}} \cup S_{\mathrm{bad}} \).
  \begin{enumerate}
  \item If \( s \) contains \( \clockfont{+1} \) as a substring (say, in positions \( k \) and \( k+1 \)), then any string in \( \cC(s) \) contains \( \clockfont{+1} \) in positions \( k \) and \( k+1 \).
    \item If \( s \) contains \( \clockfont{0-} \) in positions \( k \) and \( k+1 \), then any string in \( \cC(s) \) contains \( \clockfont{0-} \) in positions \( k \) and \( k+1 \).
  \end{enumerate}
\end{lem}
\begin{proof}
  It suffices to consider the four (two-way) rewrite rules
  \begin{align*}
    \clockfont{[0+} &\Leftrightarrow \clockfont{[1+} \\
    \clockfont{1+0} &\Leftrightarrow \clockfont{1-0} \\
    \clockfont{-0+} &\Leftrightarrow \clockfont{-1+} \\
    \clockfont{-0]} &\Leftrightarrow \clockfont{-1]}
  \end{align*}
  on string \( s \), where \( \clockfont{[} \) marks the beginning of a string and \( \clockfont{]} \) marks the end of a string.
  We first show (1).
  By the second rewrite rule, position \( k+1 \) must be \( \clockfont{0} \) in order for position \( k \) to change under a rewrite starting from \( s \).
  Likewise, by the remaining rewrite rules,
  position \( k \) must be \( \clockfont{-} \) in order for position \( k+1 \) to change under a rewrite.
  Together, this implies that positions \( k \) and \( k+1 \) are invariant under rewrites from \( s \), as desired.
  The analysis is analogous to show (2).
\end{proof}

\begin{lem}
  \label{lem:comp-onef}
  For any \( f \in S_{\mathrm{fake}} \),
  \( f \) is the only element of \( S_{\mathrm{fake}} \) which is in \( \cC(f) \).
\end{lem}
\begin{proof}
  Fix distinct \( f, g \in S_{\mathrm{fake}} \).
  We want to show that \( f \) and \( g \) are in different components of \( G \).
  We case on the the first position \( k \in [T] \) at which \( f_k \ne g_k \).
  Suppose \( k = 1 \) and without loss of generality take \( f_1 = \clockfont{0} \) and \( g_1 = \clockfont{1} \).
  Since \( f \in S_{\mathrm{fake}} \),
  we must have \( f_i = \clockfont{0} \) for all odd \( i \).

  We now argue that \( f_2 = \clockfont{-} \).
  If instead \( f_2 = \clockfont{+} \),
  then \( f_i = \clockfont{+} \) for all even \( i \),
  since \( f \in S_{\mathrm{fake}} \).
  But then \( f = \itl{\clockfont{0}^{\tau+1}}{\clockfont{+}^{\tau}} \in S_{\mathrm{good}} \),
  which is a contradiction.

  Applying \Cref{lem:lock} to \( f_1 f_2 = \clockfont{0-} \),
  every string \( h \in \cC(f) \) satisfies \( h_1 h_2 = \clockfont{0-} \).
  In particular, since \( g_1 = \clockfont{1} \), \( g_1 \notin \cC(f) \), as desired.
  We depict the above deductions as
  \begin{align*}
    f &= \underline{\clockfont{0-}}\clockfont{0*0}\tcd \\
    g &= \clockfont{1****}\tcd,
  \end{align*}
  where the underlined positions are those ``locked'' by \Cref{lem:lock},
  and \( \clockfont{*} \) indicates a position which is unconstrained (or otherwise unimportant in our argument).
  
  Next, suppose \( k = T \).
  Without loss of generality, take \( f_T = \clockfont{0} \) and \( g_T = \clockfont{1} \).
  We will deduce that
  \begin{align*}
    f &= \tcd\clockfont{****0}\\
    g &= \tcd\clockfont{1*1}\underline{\clockfont{+1}}.
  \end{align*}
  Since \( g \in S_{\mathrm{fake}}\),
  \( g_i = \clockfont{1} \) for all odd \( i \).
  Now, note that \( g_{T-1} = \clockfont{+} \):
  if instead \( g_{T-1} = \clockfont{-} \),
  then \( g_i = \clockfont{-} \) for all even \( i \),
  which puts \( g = \itl{\clockfont{1}^{\tau+1}}{\clockfont{-}^{\tau}} \in S_{\mathrm{good}} \),
  a contradiction.
  Then, applying \Cref{lem:lock} to \( g_{T-1}g_T = \clockfont{+1} \),
  every string in \( \cC(g) \) ends in \( \clockfont{+1} \),
  and thus \( f \notin \cC(g) \) since \( f_T = \clockfont{0} \).

  In the third case,
  suppose \( k \) is odd and \( 1 < k < T \).
  As usual, take \( f_k = 0 \) and \( g_k = 1 \).
  We consider two subcases corresponding to the possible values of \( g_{k-1} \).
  The case \( g_{k-1} = \clockfont{+} \) is immediate:
  \Cref{lem:lock} locks positions \( k-1 \) and \( k \) of any string in \( \cC(g) \) to \( \clockfont{+1} \),
  so \( f \notin \cC(g) \).
  In the other case \( g_{k-1} = \clockfont{-} \),
  we will deduce that
  \begin{align*}
    f &= \clockfont{1-1-}\tcd\clockfont{1-}\underline{\check{\clockfont{0}}\clockfont{-}}\clockfont{0*0}\tcd\\
    g &= \clockfont{1-1-}\tcd\clockfont{1-}\clockfont{1****}\tcd,
  \end{align*}
  where we have accented the \( k \)-th position of \( f \).
  Since \( g \in S_{\mathrm{fake}} \),
  \( g_i = \clockfont{1} \) for all odd \( i \le k\),
  and \( g_i = \clockfont{-} \) for all even \( i \le k-1 \).
  By definition, \( k \) is the first position in which \( f \) and \( g \) differ,
  so \( f_i = \clockfont{1} \) for all odd \( i < k \),
  and \( f_i = \clockfont{-} \) for all even \( i \le k-1 \).
  Moreover, since \( f \in S_{\mathrm{fake}} \),
  \( f_i = \clockfont{0} \) for all odd \( i \ge k \).
  Together, this implies that \( f_{k+1} = \clockfont{-} \);
  if instead \( f_{k+1} = \clockfont{+} \),
  then \( f_{i} = \clockfont{+} \) for all even \( i \ge k+1 \)
  and thus \( f \in S_{\mathrm{good}} \),
  which is a contradiction.
  \Cref{lem:lock} then locks positions \( k \) and \( k+1 \) of any string in \( \cC(f) \) to \( \clockfont{0-} \),
  so \( g \notin \cC(f) \), as desired.
 
  Finally, suppose \( k \) is even and take \( f_k = \clockfont{+} \), \( g_k = \clockfont{-} \).
  This case is analogous to the previous one.
  First, if \( g_{k-1} = \clockfont{0} \) then \Cref{lem:lock} locks positions \( k-1 \) and \( k \) of any string in \( \cC(g) \) to \( \clockfont{0-} \), so \( f \notin \cC(g) \).
  If instead \( g_{k-1} = \clockfont{1} \),
  then it is straightforward to check that \( f \) and \( g \) have the form
  \begin{align*}
    f &= \clockfont{1-1-}\tcd\clockfont{1}\underline{\check{\clockfont{+}}\clockfont{1}}\clockfont{+*+}\tdt\tdt\tdt\\
    g &= \clockfont{1-1-}\tcd\clockfont{1-}\clockfont{****}\tdt\tdt\tdt
  \end{align*}
  using the reasoning of the previous case.
  Thus \( g \notin \cC(f) \), completing the proof.
\end{proof}

\paragraph{Optimizing the energy over one component.}
We now bound the minimum of the quadratic form appearing in \Cref{eq:energy-components} for a component containing a fake string, and show that it is lower-bounded by a constant independent of the number of bad strings in the component. We first show the desired bound in a special case where the computational Hilbert space $\hH$ is one-dimensional, from which the general case will follow as an easy corollary.
\begin{lem}\label{lem:opt-by-hand}
    Let $k \geq 1$ be an integer and let $\psi = (\psi_0, \dots, \psi_{k}) \in \mathbb{C}^{k+1}$ be a unit vector. Then 
    \begin{align} 
    f(\psi) := \sum_{i=1}^{k}  (\frac{1}{2} |\psi_0 - \psi_i|^2 + |\psi_i|^2) \geq 1/4.
    \end{align}
\end{lem}
\begin{proof}
We would like to minimize
\begin{align} 
f(\psi) = \sum_{i=1}^{k}  (\frac{1}{2} |\psi_0 - \psi_i|^2 + |\psi_i|^2)
\end{align}
subject to $\psi$ being a unit vector in $\mathbb{C}^{k+1}$. Our first observation is to see that we can without loss of generality take all coordinates of $\psi$ to be real and nonnegative. Next, observe that
\begin{align}
    \sum_{i=1}^k |\psi_0 - \psi_i|^2 &= \sum_{i=1}^k (\psi_0^2 +  \psi_i^2 - 2 \psi_0 \psi_i) \\
    &= k \psi_0^2 + \sum_{i=1}^k (\psi_i^2) -2 \psi_0 \sum_i \psi_i \\
    &= k \psi_0^2 + (1 - \psi_0^2) - 2\psi_0 \sum_{i=1}^{k} \psi_i \\
    &\geq k \psi_0^2 + (1 - \psi_0^2) -2 \psi_0 \sqrt{k} \cdot \sqrt{\sum_{i=1}^{k} \psi_i^2} \label{eq:cs} \\
    &= k\psi_0^2 - 2\psi_0\sqrt{k} \sqrt{1 - \psi_0^2} + (1- \psi_0^2) \\
    &= (\sqrt{k} \psi_0 - \sqrt{1 - \psi_0^2})^2,
\end{align}
where we have used the Cauchy-Schwarz inequality in passing to~\eqref{eq:cs}, and then applied the normalization condition. So overall the quantity we want to minimize is
\begin{align}
    f(\psi) \geq g(\psi_0) &:= \frac{1}{2} (\sqrt{k} \psi_0 - \sqrt{1 - \psi_0^2})^2 + (1 - \psi_0^2) \\
    &= \begin{pmatrix} \psi_0 & \sqrt{1 - \psi_0^2} \end{pmatrix} \left ( \frac{1}{2} \begin{pmatrix} \sqrt{k} & -1 \end{pmatrix} \begin{pmatrix} \sqrt{k} \\ -1 \end{pmatrix} + \begin{pmatrix} 0 & 0 \\ 0 & 1 \end{pmatrix} \right) \begin{pmatrix} \psi_0 \\ \sqrt{1 - \psi_0^2} \end{pmatrix} \\
    &= \begin{pmatrix} \psi_0 & \sqrt{1 - \psi_0^2} \end{pmatrix} \cdot  \underbrace{\begin{pmatrix} k/2 & - \sqrt{k}/2 \\ -\sqrt{k}/2 & 3/2 \end{pmatrix}}_{G} \cdot \begin{pmatrix} \psi_0 \\ \sqrt{1 - \psi_0^2} \end{pmatrix}.
\end{align}
Thus, we have reduced the problem to finding the minimum eigenvalue of the $2\times 2$ matrix $G$. Through explicit computation, we can find the eigenvalues of this matrix.
\begin{align}
    \det(G - \lambda I) &= 0 \\
    (k/2 - \lambda) \cdot (3/2 - \lambda) - k/4 &= 0 \\
    \lambda^2 - \frac{3+k}{2}\lambda + \frac{3k}{4} - \frac{k}{4} &= 0 \\
    \lambda^2 - \frac{3+k}{2} \lambda + \frac{k}{2} &= 0 \\
    \lambda &= \frac{ (3+k)/2 \pm \sqrt{(3+k)^2/ 4 - 2k}}{2} \\
    &= \frac{3+k}{4} \cdot (1 \pm \sqrt{1 - 8k/(3+k)^2}).
\end{align}
Now let us bound this expression. First, assume that $k \geq 3$. Then $(3+k) \leq 2k$, so we have
\begin{align}
    \frac{8k}{(k+3)^2} &\geq \frac{8k}{(2k)^2} = \frac{2k}{k^2} \geq \frac{2k-1}{k^2}.
\end{align}
Using this bound, we get
\begin{align}
\frac{8k}{(k+3)^2} &\geq \frac{2k-1}{k^2} \\
 1 - \frac{8k}{(3+k)^2} &\leq 1 - \frac{2}{k} + \frac{1}{k^2} = 1 - \frac{2k-1}{k^2} \\  
   \sqrt{1 - \frac{8k}{(3+k)^2}} &\leq 1 - \frac{1}{k} \\
   1 - \sqrt{1 - \frac{8k}{(3+k)^2}} &\geq 1/k \\  
\end{align}
So we have, for $k \geq 3$, the bound
\[ \lambda \geq \frac{3+k}{4} \cdot \frac{1}{k} \geq \frac{1}{4}. \]
For $k \in \{1,2\}$, we can check that by explicit computation, $\lambda \geq 1/4$. (In fact, the ``true'' lower bound for all $k$ appears to be $1 - 1/\sqrt{2} \approx 0.293$, which is attained for $k = 1$, but the looser bound of $1/4$ is good enough for us.)
\end{proof}

\begin{cor}\label{cor:opt-by-hand-with-u}
Let $k, d \geq 1$ be integers and let $\ket{\alpha_0}, \dots, \ket{\alpha_k}$ be (not necessarily unit) vectors over $\mathbb{C}^d$ such that $\sum_{i=0}^{k} \| \ket{\alpha_i} \|^2 = 1$. Moreover, let $U_1, \dots, U_k$ be $d \times d$ unitary matrices. Then 
    \begin{align} 
    \sum_{i=1}^{k}  (\frac{1}{2} \| \ket{\alpha_0} - U_i \ket{\alpha_i} \|^2 + \|\ket{\alpha_i} \|^2) \geq 1/4.
    \end{align}    
\end{cor}
\begin{proof}
    Observe that for any two vectors $\ket{\alpha}, \ket{\beta}$ and for any unitary $U$, it holds that
\begin{align}
    \| \ket{\alpha} - U \ket{\beta} \|^2 &= \| \ket{\alpha} \|^2 + \| \ket{\beta} \|^2 - \bra{\alpha} U \ket{\beta} - \bra{\beta} U \ket{\alpha} \\
    &\geq \| \ket{\alpha} \|^2 + \| \ket{\beta} \|^2 - 2 \| \ket{\alpha} \| \cdot \| \ket{\beta} \| \\
    &= | \| \ket{\alpha} \| - \| \ket{\beta} \| |^2. \label{eq:align-the-vectors}
\end{align}
Thus, if we let $\psi_i = \| \ket{\alpha_i} \|$, we see that the vector $(\psi_0, \dots, \psi_k)$ satisfies the conditions of \Cref{lem:opt-by-hand}, and by \Cref{eq:align-the-vectors}, the quantity we wish to bound is at least $f(\psi) \geq 1/4$ by \Cref{lem:opt-by-hand}.
\end{proof}

\bibliographystyle{myhalpha}
\bibliography{editedxzhambib,anand}

\appendix

\section{Commuting $\XZQkSAT$ is in $\NP$}
The special case of our $\XZQkSAT$ where all terms are promised to commute is contained in $\NP$ (and in fact $\NP$-complete). This result is folklore but we include a simple self-contained proof here. It uses a simple version of the so-called ``trace technique,'' which was used by Schuch~\cite{Schuch11} to analyze commuting Hamiltonians on lattices.
\begin{thm}\label{thm:commuting-xz-in-np}
    The problem of, given an instance of $\XZQkSAT$ where it is promised that all terms in the Hamiltonian commute, deciding whether the ground energy is $0$ or $> 0$ is $\NP$-complete.
\end{thm}
To prove this theorem, we will need two auxiliary lemmas. First, the following lemma characterizes the projector onto the zero-energy eigenspace of a commuting Hamiltonian with projector terms.
\begin{lem} \label{lem:proj}
  Let \( H = \sum_j h_j \) be a commuting Hamiltonian with projector terms.
  Then the projector onto the zero-energy eigenspace of \( H \) can be written as
  \begin{align}
  \label{eq:27}
  \prod_{j} (I - h_j).
  \end{align}
\end{lem}
\begin{proof}
  Since the \( h_j \)'s commute (and are diagonalizable), they have a simultaneous eigenbasis, which we denote by \(B = \set{\ket{\varphi_i}}_i \).
  The projector onto the zero-energy eigenspace of \( H \) is, by definition, the linear operator \( P \) which has the following action on the basis \( B \):
  \( P \ket{\varphi_i} = \ket{\varphi_i}\) if \( H \ket{\varphi_i} = 0 \), and \( P \ket{\varphi_i} = 0  \) if \( H \ket{\varphi_i} \ne 0 \).

  We show that \( \prod_{j} (I - h_j) = P\) by showing that it acts the same on \( B \).
  Suppose \( H \ket{\varphi_i} = 0 \).
  Then \( h_j \ket{\varphi_i} = 0 \) for all \( j \),
  so \( \prod_{j} (I - h_j) \ket{\varphi_i} = \ket{\varphi_i} \).

  On the other hand, if \( H \ket{\varphi_i} \ne 0 \), then there is some \( k \) such that \( h_k \ket{\varphi_i} \ne 0\).
  But \( \ket{\varphi_i} \) is an eigenstate of \( h_k \), and it can only have eigenvalue \( 0 \) or \( 1 \) since \( h_k \) is a projector.
  Thus, it is a \( 1 \)-eigenstate, and \( h_k \ket{\varphi_i} = \ket{\varphi_i} \).
  Then \( (I - h_k) \ket{\varphi_i} = 0 \), so \( \prod_{j}(I - h_j) \ket{\varphi_i} = 0 \) by commuting \( I - h_k \) to the right.
  This completes the proof.
\end{proof}

The second lemma relates the existence of a zero-energy state for a Hamiltonian $H$ to the trace of a projector.
\begin{lem}
  \label{lem:trace}
  Let \( H \) be a Hamiltonian and let \( P \) be the projector onto the zero-energy eigenspace of \( H \).
  Then \( \tr P \ne 0\) if and only if \( H \) has a zero-energy state. 
\end{lem}
\begin{proof}
  \( H \) has a zero-energy state if and only if the dimension of the zero-energy eigenspace of \( H \) is not \( 0 \).
  But \( \tr P \) calculates the sum of the eigenvalues of \( P \),
  which is just the dimension of the zero-energy eigenspace of \( H \),
  since \( P \) is a projector.
\end{proof}

\begin{proof}[Proof of \Cref{thm:commuting-xz-in-np}]
    Firstly, observe that $\NP$-hardness is straightforward as any instance of $\lang{3-SAT}$ is automatically an instance of $\XZQkSAT$. So all that remains is to show that this problem is contained in $\NP$.
    
    Given an instance $H$ of $\XZQkSAT$ with commuting terms on $n$ qubits, write it as
    \[H = H_X + H_Z,\]
    where $H_X$ gathers all terms that are diagonal in the $X$-basis and $H_Z$ all terms that are diagonal in the $Z$-basis. We claim that the following $\NP$ protocol is complete and sound for this problem: the verifier receives a pair of $n$-bit strings $x, z$, and accepts if $\bra{z}H_Z \ket{z} = 0$ and $\bra{x} \had^{\otimes n} H_X \had^{\otimes n} \ket{x} = 0$, and rejects otherwise. (This verifier runs in polynomial time since it is possible to check in polynomial time for each individual term in $H_Z$ whether it annihilates a standard basis state $\ket{z}$, and for each individual term in $H_X$ whether it annihilates a Hadamard basis state $\had^{\otimes n} \ket{x}$.)

    Firstly, for completeness, suppose $H$ has a ground state $\ket{\psi}$ with ground energy $0$. Then it follows that for every string $z$ in the support of $\ket{\psi}$ in the standard basis, $\bra{z}H_Z\ket{z} =0$, and for every string $x$ in the support of $\ket{\psi}$ in the Hadamard basis, $\bra{x}\had^{\otimes n} H_X \had^{\otimes n} \ket{x} = 0$. So an accepting pair of string $x,z$ necessarily exists.

    For soundness, let us introduce notation for the individual terms of the Hamiltonian. Write
    \begin{align*}
        H_X &= \sum_{i} h_{X,i} \\
        H_Z &= \sum_{j} h_{Z,j},
    \end{align*}
    where each operator $h_{X,i}, h_{Z,j}$ is a projector. 
    Let
    $\Pi$ be the projector onto the $0$-energy space of $H$, $\Pi_X$ be the projector onto the $0$-energy space of $H_X$, and $\Pi_Z$ be the projector onto the $0$-energy space of $H_Z$. By \Cref{lem:proj}, we see that
    \begin{align*}
        \Pi_X &= \prod_i (I - h_{X,i}) \\
        \Pi_Z &= \prod_j (I - h_{Z,j}) \\
        \Pi &= \Pi_X \Pi_Z = \Pi_Z \Pi_X.
    \end{align*}
    Now, suppose $x^*,z^*$ is an accepting pair of strings in our $\NP$ protocol. Then we claim that $\tr \Pi > 0$, and thus by \Cref{lem:trace}, $H$ has a zero-energy ground state. Indeed, define the set of ``good'' $Z$ strings $G_Z = \{z \in \{0,1\}^n: H_Z \ket{z} = 0\}$ and $G_X = \{x \in \{0,1\}^n: H_X \had^{\otimes n} \ket{x} = 0\}$. Then we can write
    \begin{align*}
        \Pi_X &= \sum_{x \in G_X} \had^{\otimes n} \ket{x}\bra{x} \had^{\otimes n} \\
        \Pi_Z &= \sum_{z \in G_Z} \ket{z}\bra{z}.
    \end{align*}
    Moreover, the existence of $x^*, z^*$ implies that the sets $G_X, G_Z$ are nonempty. Now, we compute the trace:
    \begin{align}
        \tr(\Pi) &= \tr(\Pi_X \Pi_Z) \\
        &= \sum_{x \in G_X} \sum_{z \in G_Z} \tr( \had{\otimes n} \ket{x}\bra{x} \had^{\otimes n} \ket{z}\bra{z} ) \\
        &= \sum_{x \in G_X} \sum_{z \in G_Z} |\bra{x} \had^{\otimes n} \ket{z} |^2 \\
        &\geq |\bra{x^*} \had^{\otimes n} \ket{z^*}|^2 \\
        &> 0,
    \end{align}
    where in the last line we have used the fact that every standard basis state and every Hadamard basis state have non-zero overlap.
\end{proof}

\end{document}